# Synthesis of hard magnetic α-MnBi phase by high pressure torsion and field assisted annealing

Lukas Weissitsch [a,*], Stefan Wurster [a], Martin Stückler [a,§], Timo Müller [b,†], Heinz Krenn [c], Reinhard Pippan [a], Andrea Bachmaier [a]

[a] *Erich Schmid Institute of Materials Science of the Austrian Academy of Sciences, 8700 Leoben, Austria*
[b] *Deutsches Elektronen Synchrotron (DESY), 22607 Hamburg, Germany*
[c] *Institute of Physics, University of Graz, 8010 Graz, Austria*
[§] *Current address: TDK Electronics GmbH & Co OG, 8530 Deutschlandsberg, Austria*
[†] *Current address: Anton Paar GmbH, Anton-Paar-Str. 20, 8054 Graz, Austria*
[*] *Corresponding author: E-Mail address: lukas.weissitsch@oeaw.ac.at*

**Graphical Abstract**

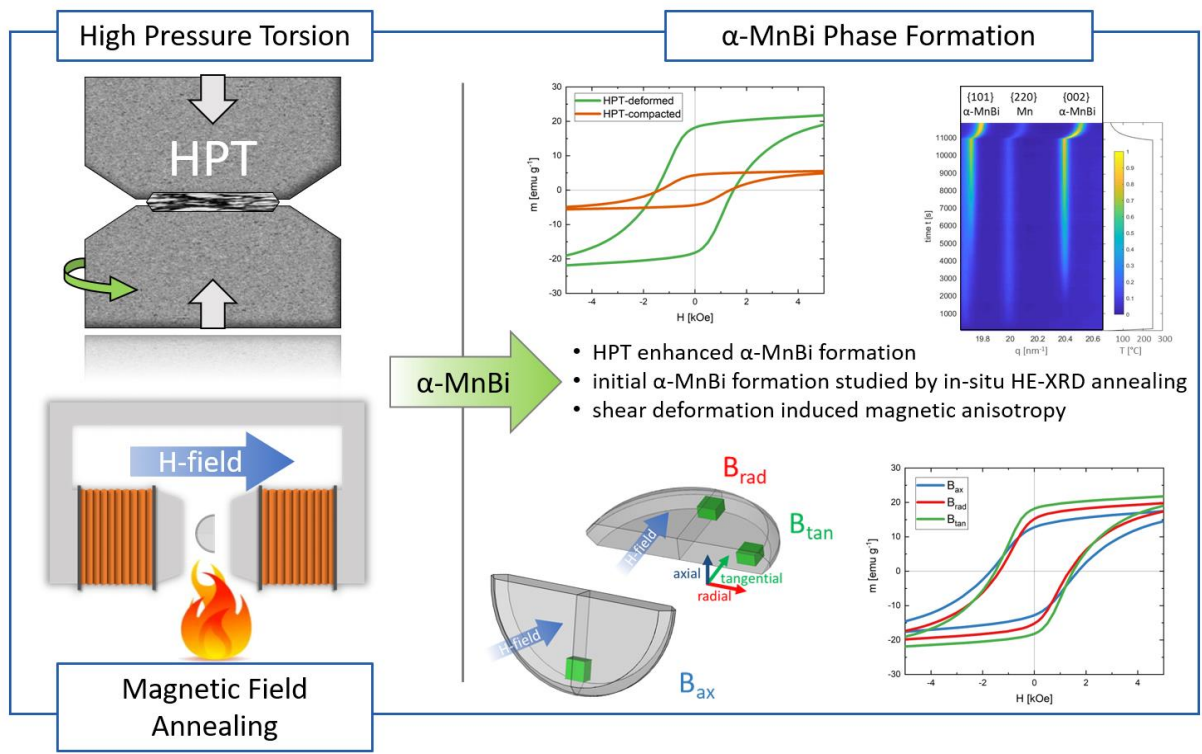

**Abstract**

Bulk composites using powder blends of Mn and Bi with equal atomic ratios are consolidated and severely deformed by high-pressure torsion (HPT). Subsequent annealing treatments lead to the formation of the ferromagnetic and rare-earth free hard magnetic α-MnBi phase. The initial phase formation is studied by in-situ high-energy XRD annealing experiments showing a beneficial influence of HPT-deformation on the amount of α-MnBi. The nucleation sites are strongly increased, thus the volumetric α-MnBi phase content exceeds 50 vol.%. A lower HPT-pressure of 2 GPa is found to be preferred over 5 GPa to obtain high α-MnBi contents. Annealing is done with or without applying an external magnetic field. Additionally, influences of the HPT-induced shear deformation are discussed and correlated with preferred diffusion path ways fostering the formation of anisotropic α-MnBi phase.

**Introduction**

For several decades, the MnBi system has attracted attention of researches. Already since the 1950's, the magnetic properties of this material system have been studied, fostering the possible application as permanent magnet material [1,2], and the equilibrium phase diagram has been constantly improved [1–5]. Nowadays, phase transitions and their crystal structures are widely understood [6,7]. The recent research interest focuses on the intermetallic and ferromagnetic low temperature phase of equiatomic composition, herein denoted as ´α-MnBi phase`, as it exhibits exceptional hard magnetic properties. A rare and very interesting phenomenon is the positive temperature coefficient of intrinsic coercivity for temperatures T < 500 K [8,9], originating in a large and peaking magnetocrystalline anisotropy of about 2.2 MJ/m$^3$ above 400 K [6,10,11]. Thus, this material is particularly interesting as permanent magnet for high-temperature applications. However, two drawbacks are known: First, the low saturation magnetization ($M_s$) of hard magnetic materials, which is about 79 emu/g for α-MnBi [12]. This could be circumvented if exchange coupling between the hard magnetic α-MnBi phase and a soft magnetic material is induced. Second, processing of the α-MnBi phase in sufficient quantities is a challenging task. Thus, the main goal of this work is to show that a comparatively easy process of severe plastic deformation by high-pressure torsion (HPT), in concatenation with an adequate, subsequent annealing step, can be used to process large quantities of α-MnBi.

The current roadmap of fabricating MnBi based permanent magnets starts with the formation of a high purity α-MnBi phase in the form of an alloy or powder, which than has to be reprocessed to form bulk materials usually referred to a top-down approach. A very effective way is to utilize rapid cooling processes and subsequent annealing procedures. Several studies show the presence of α-MnBi phase in materials after arc-melting [8,13,14] or melt spinning [15–17], which then are grinded, milled, magnetically separated and compacted. Maintaining good magnetic properties during such consolidation processes is challenging and a task on its own [18]. Sometimes this includes alignment of the powders in a magnetic field or the alloying of additional elements, to obtain a dense (anisotropic) green body [19,20]. At this stage of processing, materials possess strongly refined particle sizes within the range down to the single domain size (about 250 nm [21]). A subsequent proper annealing treatment allows the formation of the high temperature β-MnBi phase, which is transferred upon slow cooling and a peritectic transition, to the α-MnBi phase [4,11,15,22,23]. Such procedure significantly increases the amount of the desired phase, already showing incredible results up to 98 wt.% α-MnBi content. However, for industrial utilization a high demand for a simplified processing route exists.

For this, a bottom-up technique is implemented, where the starting materials can be powders and one directly obtains a bulk material. To generate the desired α-MnBi phase, HPT, a severe plastic deformation method is introduced herein. HPT-deformation allows to obtain large sample sizes with dimension of several millimeters up to centimeters, whereas the microstructure simultaneously features particle and grain sizes in the range between micrometers to tens of nanometers [24]. As starting materials, bulk materials or any powder blends, which can be also consolidated in inert gas atmosphere, are convenient. To compare this technique with other deformation processes, the applied shear strain $\gamma$ is often converted to the von Mises equivalent strain $\varepsilon$, which is based on the equivalence of deformation energy and described by [25]

$$\varepsilon = \frac{\gamma}{\sqrt{3}} = \frac{1}{\sqrt{3}} \cdot \frac{2\pi r}{t} n$$

where $r$ is the radial distance of the analyzed part of the sample to the HPT rotation axis, $t$ the thickness of the cylindrical specimen in axial direction and $n$ the number of applied revolutions. Very large amounts of deformation can easily be applied, as the shear strain is directly proportional to the number of revolutions. HPT usually leads to a shear texture and a pronounced grain refinement, yet in principle, a broad variety of microstructural morphologies can be obtained [26]. The possibility to combine completely different processing steps, inter alia, powder compaction (equivalent to sintering), grain refinement and texturing is notable and will open new opportunities in the production of MnBi permanent magnets. The main drawback of HPT is the limited sample size although an upscaling is possible [27]. However, HPT is the easiest way to study the influence of high applied strains, and the results can be adapted for more continuous and industrial relevant manufacturing designs [28].

In this study, we first discuss the influence of HPT-processing parameters (e.g., applied pressure, amount of applied strain) on the formation of the α-MnBi phase during subsequent annealing. The α-MnBi phase formation is monitored by scanning electron microscopy (SEM) as well as by high energy X-ray diffraction (HEXRD) in-situ annealing experiments using synchrotron-radiation. In the second part, the annealing procedure is further improved applying a vacuum and magnetic field environment while annealing. The influence of the direction of applied magnetic field with respect to the HPT-induced microstructural texture, is extensively studied by SQUID magnetometry. The thorough investigation of the large parameter space provides a sound knowledge on influences of processing steps on the α-MnBi phase formation which further allows to set the magnetic properties by tuning the process accordingly.

**Experimental**

*Sample Preparation*

Binary powder blends with an equal atomic ratio are mixed using conventional high purity powders (Mn: Alfa Aesar 99.95% - 325 mesh; Bi: Alfa Aesar 99.999% - 200 mesh). Ball milling of the powder blends is done by using an air-cooled planetary ball mill (Retsch PM400), with a 1:20 powder-to-ball ratio at 300 rpm and a total milling time up to 4 h. To prevent heat development inside the jar, the milling process is interrupted every 60 minutes for at least 30 minutes. To prevent oxidation, the powder handling, ball-milling as well as the consolidation process is carried out in an Ar-atmosphere. The powder blends are hydrostatically consolidated in the HPT device; these samples will herein be denoted as HPT-compacted. Further samples are subsequently HPT-deformed at a nominal pressure of 2 or 5 GPa for 30 revolutions and at room temperature (RT) using a rotation speed of ~0.6 rpm supported by a high-pressure air cooling to avoid a temperature increase through inner friction. In the second part of this study, the development of an even simpler process is intended, utilizing powder blends without previous ball-milling. The powders are HPT-compacted and deformed by applying 10 revolutions at a nominal pressure of 2 GPa at RT by 0.6 rpm and high-pressure air cooling. To study the α-MnBi phase formation with respect to the HPT-induced textured microstructure of Bi, an additional Bi sample is processed from Bi-granules (HMW Hauner 99.99% < 5 mm) and deformed by 25 revolutions at a nominal pressure of 2 GPa.

*Annealing Treatments*

Quarters or eighths of HPT-deformed samples are conventionally annealed (CA) under ambient conditions at 230°C with dwell times between 4 h and 120 h. The specimens are wrapped in a protective foil before the isothermal treatment. After the annealing process, they are quenched in ethanol.

For in-situ HEXRD experiments, quarters of HPT samples are investigated by heating them with an in-situ heating stage (THMS600 Linkam, Tadworth, United Kingdom), entailing a custom-made sample holder to mount the samples inside an Ar-flushed chamber, suppressing the formation of oxides. The specimens are heated with a heating rate of 2 °Cs$^{-1}$, isothermally held at 240 °C for 3 h followed by furnace cooling down below 40 °C. The lattice parameter change of a reference Cu specimen upon the same annealing treatment settings is used to calibrate the recorded temperature measured by a thermocouple mounted in the heating plate underneath the sample.

The Magnetic field assisted Vacuum Annealing (MVA) procedure is realized by a custom-made vacuum chamber fitting into the homogeneous field region provided by the conical pole pieces (diameter of 176 mm) of an electromagnet (Type B-E 30, Bruker), operated at a constant magnetic field of 2 T. Within the vacuum chamber, two copper blocks, each containing a cartridge heating element (Keller, Ihne & Tesch Ges.m.b.H.; HHP, Ø10 x L50 mm, 100 W) are positioned. These blocks embrace an especially designed copper specimen holder. Figure 1a shows a photograph of parts of the specimen holder on the upper left and a schematic representation of the described vacuum set-up. The specimen holder hosts up to eight halved HPT discs (drawn in green), which are heated by the direct heat transfer through the copper stock. A thermocouple is mounted next to the specimens allowing to monitor the constancy of the temperature, which is set to 240°C for all MVA experiments. The applied external magnetic field is recorded using a Hall-probe (Model 475 DSP, Lakeshore), placed next to the vacuum chamber.

The applied shear strain during HPT-deformation is known to induce a crystallographic alignment (texture), which accordingly influences the magnetic properties [29]. Therefore, the applied field direction during MVA with respect to the HPT-disc orientations is expected to have an impact on the α-MnBi formation. For this reason, the sample holder allows a perpendicular and parallel mount of the sample axis in regard of the applied magnetic field. As depicted in the schematics (Figure 1b), the applied external field (H-field) penetrates the HPT-disc, allowing to extract pieces, which are field annealed parallel to their axial, tangential or radial HPT-disc direction. In the following, these samples are denoted as $B_{ax}$, $B_{tan}$ and $B_{rad}$, respectively.

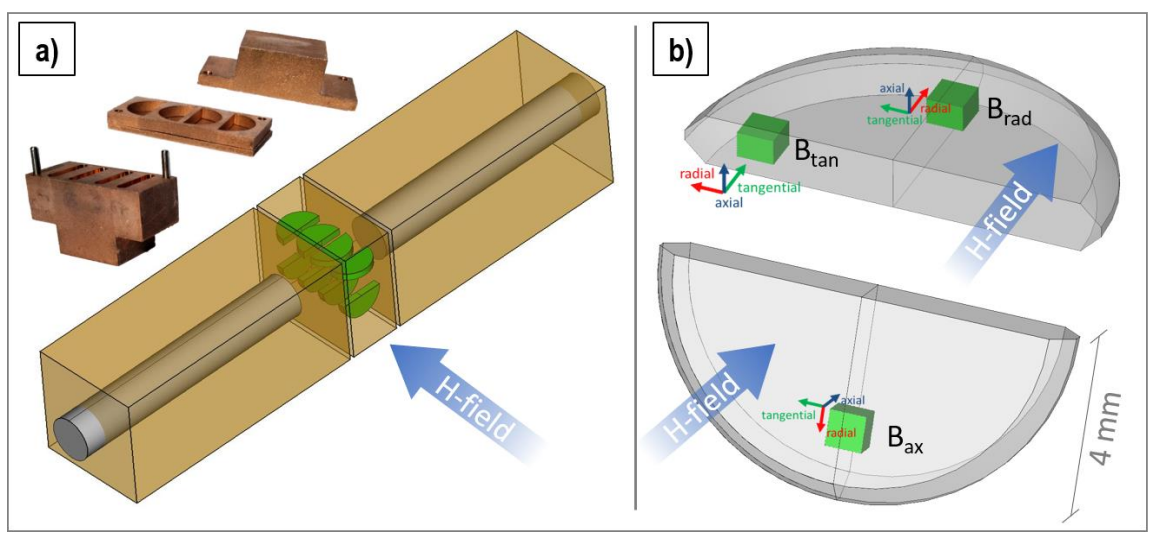

*Figure 1: a) Schematic representation of the vacuum annealing construction. Two heating elements are embedded in copper blocks, heating the sample holder (photograph on upper left) by direct contact. The samples (illustrated in green) can be mounted in two different orientations (parallel and perpendicular to the applied external field). b) The external applied field penetrates the sample in three different directions with respect to the HPT-disc orientation (radial, tangential, axial). Within one sample, three pieces for SQUID measurements with different field annealing directions are obtained.*

*Microstructural Characterization*

The microstructure is investigated using a SEM (LEO 1525, Carl Zeiss Microscopy GmbH), where images are recorded in backscattered electron (BSE) mode in tangential HPT-disc orientation. Electron backscatter diffraction (EBSD) examination is performed with a Bruker e⁻-Flash[FS] detector and for chemical analysis, energy dispersive X-ray spectroscopy (EDX; XFlash 6|60 device, Bruker) is applied. X-ray diffraction measurements (XRD; Bruker Phaser D2, Co-$K_\alpha$-radiation) are applied to the plane surface with the diffraction vector in the axial direction of the HPT-disc. References for expected peak positions are taken from the Crystallography Open Database (Mn: COD 9011108, Bi: COD 9008576, α-MnBi: COD 9008899) [30]. Hardness measurements (Micromet 5104, Buehler) are performed in tangential HPT-disc direction on the polished sample with an indention load of 50 – 100 g. The sample's density is determined by a balance (Sartorius Secura225D-1S) and a density determination kit (Sartorius YDK03), applying the principle of Archimedes. The density of the reference medium (soapy water, ρ = 0.997105 g/cm³) is measured using a reference volume (10 cm³). Magnetic results are obtained by a SQUID-magnetometer (Quantum Design MPMS-XL-7, Quantum Design, Inc.,

San Diego, CA, USA) operated with the manufacturer's software MPMSMultiVu Application (version 1.54). All hysteresis loops are measured between ±65 kOe. Synchrotron high-energy X-ray diffraction (HEXRD) experiments are performed in transmission mode, parallel to the axial HPT-disc direction, at the beamline P21.2 at PETRA III (DESY, Hamburg, Germany) [31]. The spot size is set to 200 × 200 $\mu m^2$ and a photon energy of 60 keV is used. Wide-angle X-ray scattering (WAXS) diffraction patterns are recorded by two Varex XRD 4343 CT flat panel detectors at 0.2 Hz during the in-situ annealing treatment. The experimental setup is calibrated using $LaB_6$ and 2D patterns are azimuthally integrated using PyFAI [32].

**Results**

*Influence of pressure and HPT-deformation on α-MnBi formation*

A highly uniform Mn and Bi phase distribution is required to focus on the characteristics of the α-MnBi phase formation, which is realized by ball milling of the powder blends. To monitor the ball milling process, the milling process is interrupted every 60 minutes and measured by XRD. The XRD pattern are shown in Figure 2 wherein the position for theoretical reference peaks of Mn, Bi and the α-MnBi phases are indicated by vertical lines. The diffraction patterns reveal peak broadening within the first two hours but no further significant changes regarding phase evolution, peak intensity changes or peak broadening up to 4 hours milling time are visible. Thus, after 4 hours a steady state seems to be well established and this powder is used for further experiments.

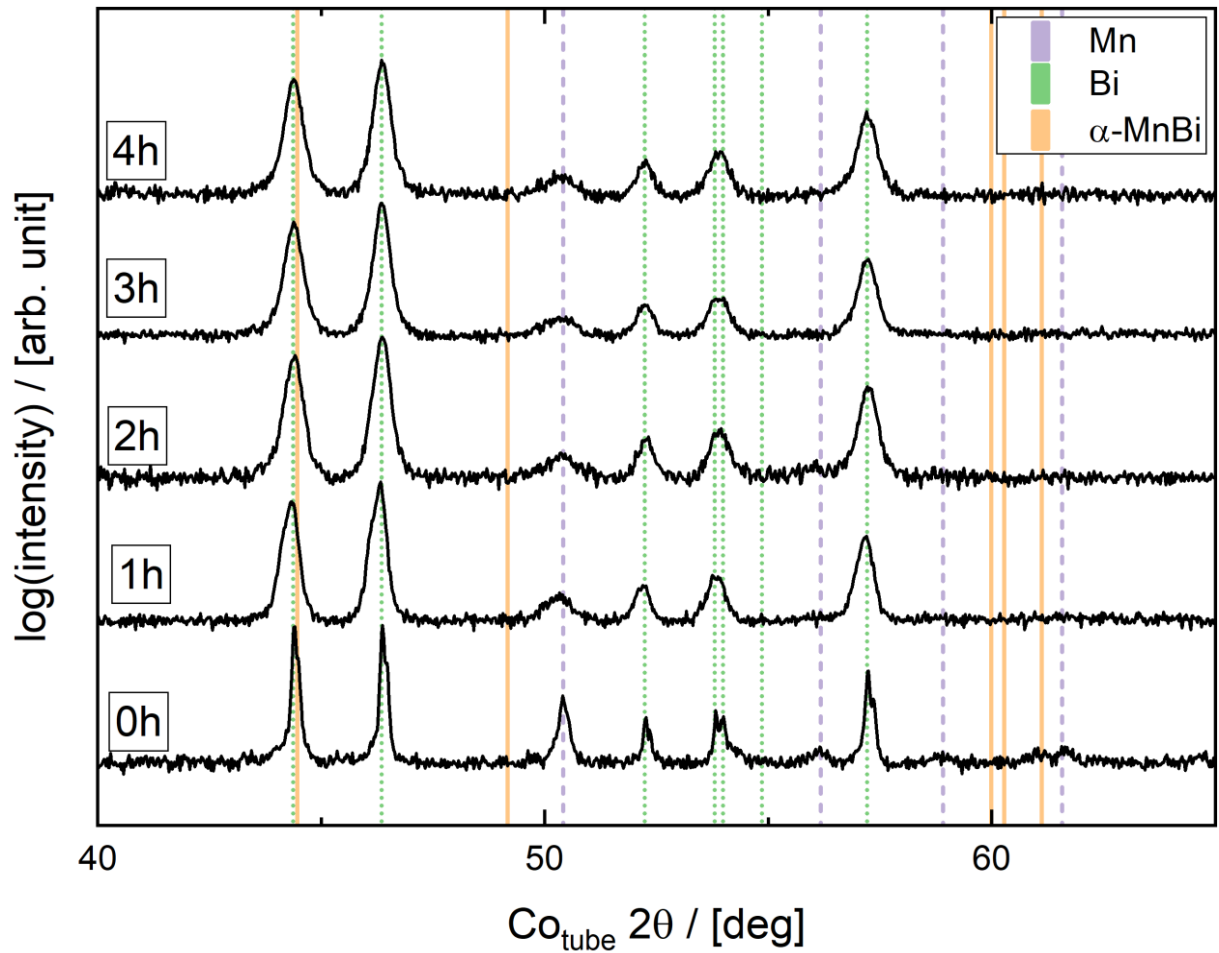

*Figure 2: XRD measurements of a MnBi powder blend after ball milling for different times. Peak broadening stagnates after 2 h of milling time. No α-MnBi phase formation is detected.*

To separate the influence of HPT-consolidation and HPT-deformation on the α-MnBi phase formation, further experiments are performed on compacted powder blends, whereby one sample is compacted at 2 GPa and another one at 5 GPa. The compacted samples are cut and suspended to annealing treatments at 230°C in ambient conditions for different times. BSE images recorded in tangential HPT-disc direction are presented in Figure 3. The samples compacted at 2 GPa and 5 GPa are shown in the first and second column respectively, while the annealing time increases from top to bottom. In the as-compacted state (labelled with 0 h) a microstructure with homogeneously dispersed Mn particles, is found. After 4 h annealing, a contrast change of the 2 GPa compacted sample indicates the formation of a new phase (marked by arrows).

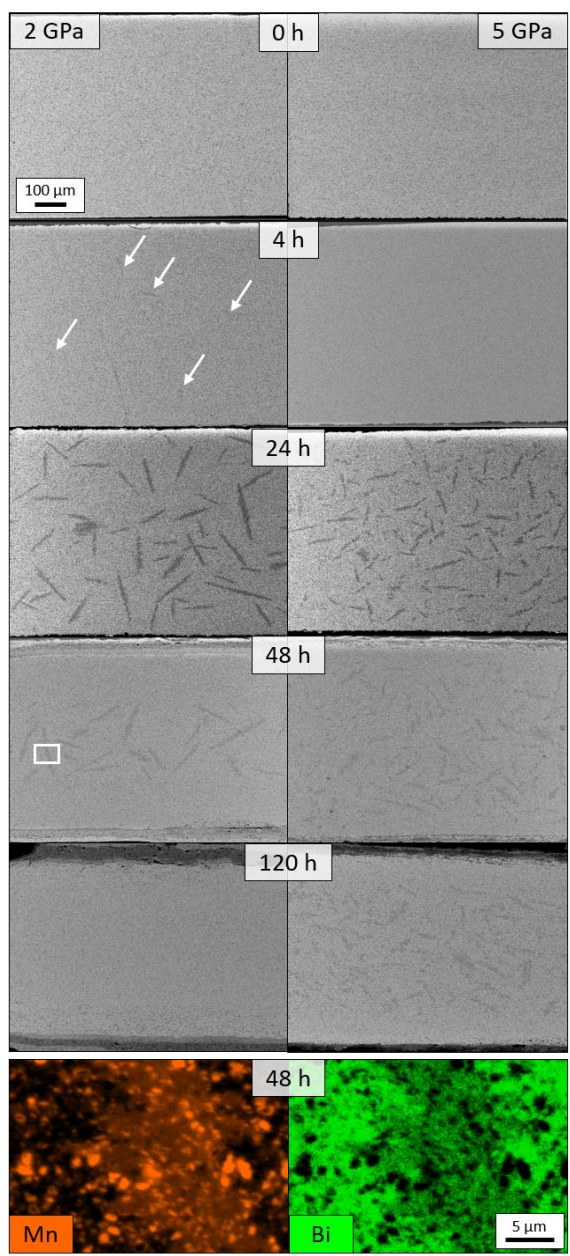

*Figure 3: BSE images of HPT- compacted MnBi powder blends, compacted at 2 GPa or 5 GPa (left and right column) and annealed under ambient conditions for different times at 230 °C. The last row shows a magnified region of the 2 GPa and 48 h annealed state (white-rimed rectangle), where the distribution of Mn and Bi is visualized by the EDX signal. The cross-sections are recorded in tangential HPT-disc direction, with the axial direction pointing vertically.*

When annealing for 24 h, this phase appears as cigar-shaped silhouettes for both samples. However, larger, but less "cigar-regions" are found for the 2 GPa sample compared to the 5 GPa sample. EDX mapping of a part of one of these "cigar-regions" (shown in the last row of Figure 3) reveal an equally strong and coeval appearance of Mn (orange) and Bi (green). Quantitative analysis gives an indication for the presence of the α-MnBi phase, but small Mn particles being present within the analyzed volume impede accurate quantification. The uniform distribution of Mn and Bi does not prevail outside these "cigar-regions", where a strict distinction between the two phases is present. In addition, indentation hardness also revealed differences between the new forming phase (66 ± 3 HV) and the surrounding matrix (40 ± 2 HV).

Unexpectedly, the α-MnBi phase starts to vanish after annealing for 48 h. This behavior is stronger pronounced for the 2 GPa sample and in addition, the morphology of the sample surface changes, as an oxide layer is formed. Confirmed by EDX measurements (see Figure Supplementary 1), mainly Mn-oxide formation occurs and the development of this layer inversely scales with the visible amount of α-MnBi phase.

To investigate the α-MnBi phase formation in more detail, in-situ annealing HEXRD experiments are conducted for samples processed with different applied pressure (2 GPa and 5 GPa) and strain (HPT-compacted and HPT-deformed to 30 turns). Integrated intensity plots of every measurement during the complete annealing experiment for all four samples are shown in Figure 4.

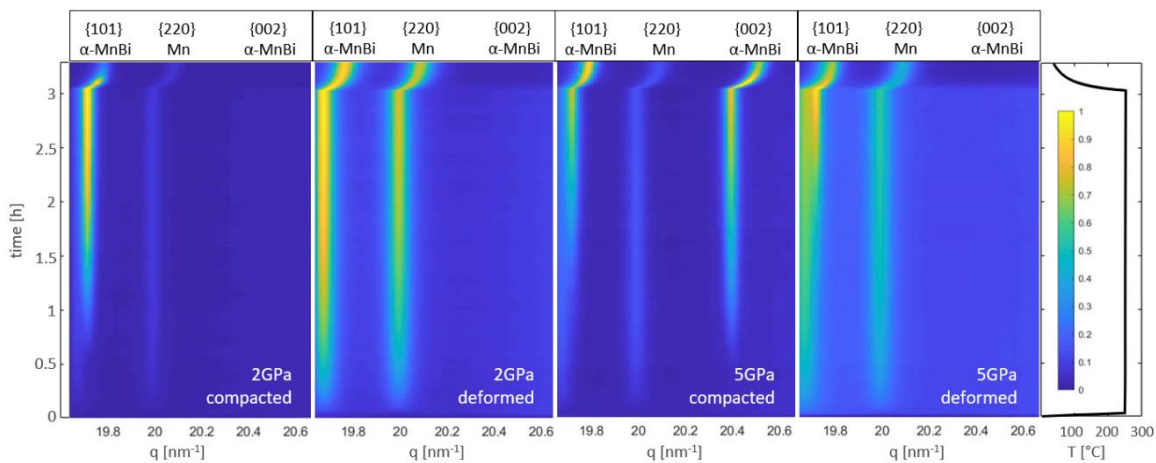

*Figure 4: Intensity plots of in-situ HEXRD annealing experiments. Four samples, compacted and deformed by 30 revolutions at 2 and 5 GPa are depicted. On the right, the corresponding temperature profile and an intensity scale bar are plotted. The peak intensities are normalized within the presented figures to the respective most intense peak. The q range is limited to a small representative fraction of the measured q-range, which includes peaks from the relevant phases, to facilitate visual comparison.*

The {220} Mn peak is found in all recorded spectra and the α-MnBi phase forms in each performed experiment. For all samples, the {101} α-MnBi peak is visible at the end of the annealing treatment but arises after different times. After about 200 s, the {101} α-MnBi peak is visible for both HPT-deformed samples. For the compacted samples, the same peak requires more time to emerge and arises after roughly 2000 s and 5000 s for 2 GPa and 5 GPa, respectively. This is more difficult to identify for the 5 GPa compacted sample, as an additional peak appears at a slightly lower q-value, which is most likely dedicated to the formation of an oxide which forms faster than the {101} α-MnBi peak. An additional {002} α-MnBi peak is only visible for the 5 GPa compacted sample.

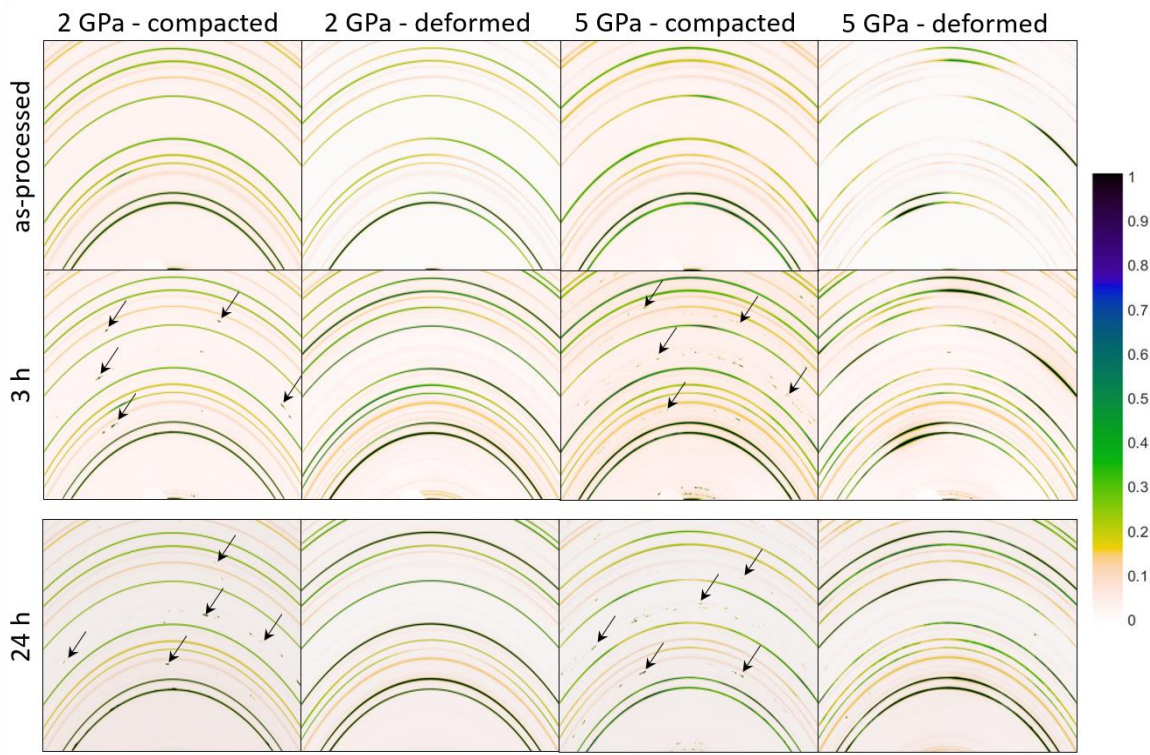

*Figure 5: Selected area detector diffraction patterns of the HEXRD annealing experiments. Tiny spots with high intensities (arrows) are visible for the HPT-compacted samples, while continuous lines with low intensities are found at the same diffraction angle for the HPT-deformed samples. Both of them correspond to the forming α-MnBi phase. The color scheme referes to the normalized intensity of the diffraction pattern.*

To better understand the observed intensities, sections of the first (initial state = as-processed sample; first row) as well as the last diffractogram of the in-situ annealing experiment (annealed for 3 h; second row) are shown in Figure 5. Due to the chosen experimental setup, only a section of the Debye-Scherrer rings is recorded. For the compacted samples, bright spots (exemplarily marked by arrows) indicate the α-MnBi phase formation. For the deformed samples continuous lines with lower intensities at the same diffraction angles confirm the existence of the α-MnBi phase. Diffraction images recorded after 24 h annealing (last row in Figure 5), which was performed ex-situ, again show bright spots for compacted samples and continuous rings with low intensity for deformed samples. These differences prove a larger number of differently oriented α-MnBi crystallites in the HPT deformed samples. Furthermore, crystallographic texture is present in the as-processed samples (in particular after HPT-deformation has been applied) as can be seen by the non-homogeneous intensity distribution along the Debye-Scherrer rings.

*Influence of magnetic field assisted annealing on the α-MnBi phase formation*

Processing of the α-MnBi phase is further improved by MVA, and simplified by using powder blends without ball-milling and a maximum HPT-pressure of 2 GPa. 2 GPa was chosen, as α-MnBi phase formation is faster in this case according to the in-situ HEXRD annealing results. Figure 6 shows hysteresis loops of an HPT-compacted and HPT-deformed sample after MVA at 2 T for 4h, both annealed tangentially ($B_{tan}$ c.f. Figure 1b). The samples are measured between ±65 kOe at 300 K. Values for $M_s$ are acquired by linearly extrapolating the magnetization above 35 kOe against $H^{-1}$ and determining the intercept at $M(H^{-1}) = 0$ [33,34]. For the compacted and MVA-treated sample (triangle, brown curve), a $M_s$ of 9.89 emu/g is found. The deformed sample shows a $M_s$ of 33.92 emu/g at 300 K measurement temperature (cube, green curve) and a $M_s$ of 36.12 emu/g at 10 K (empty cube, blue curve). The coercivity $H_c$ slightly increases from 1.38 kOe to 1.58 kOe for HPT-compacted and HPT-deformed samples, respectively. The $H_c$ of the latter one significantly decreases by one order of magnitude to 0.16 kOe, if measured at 10 K.

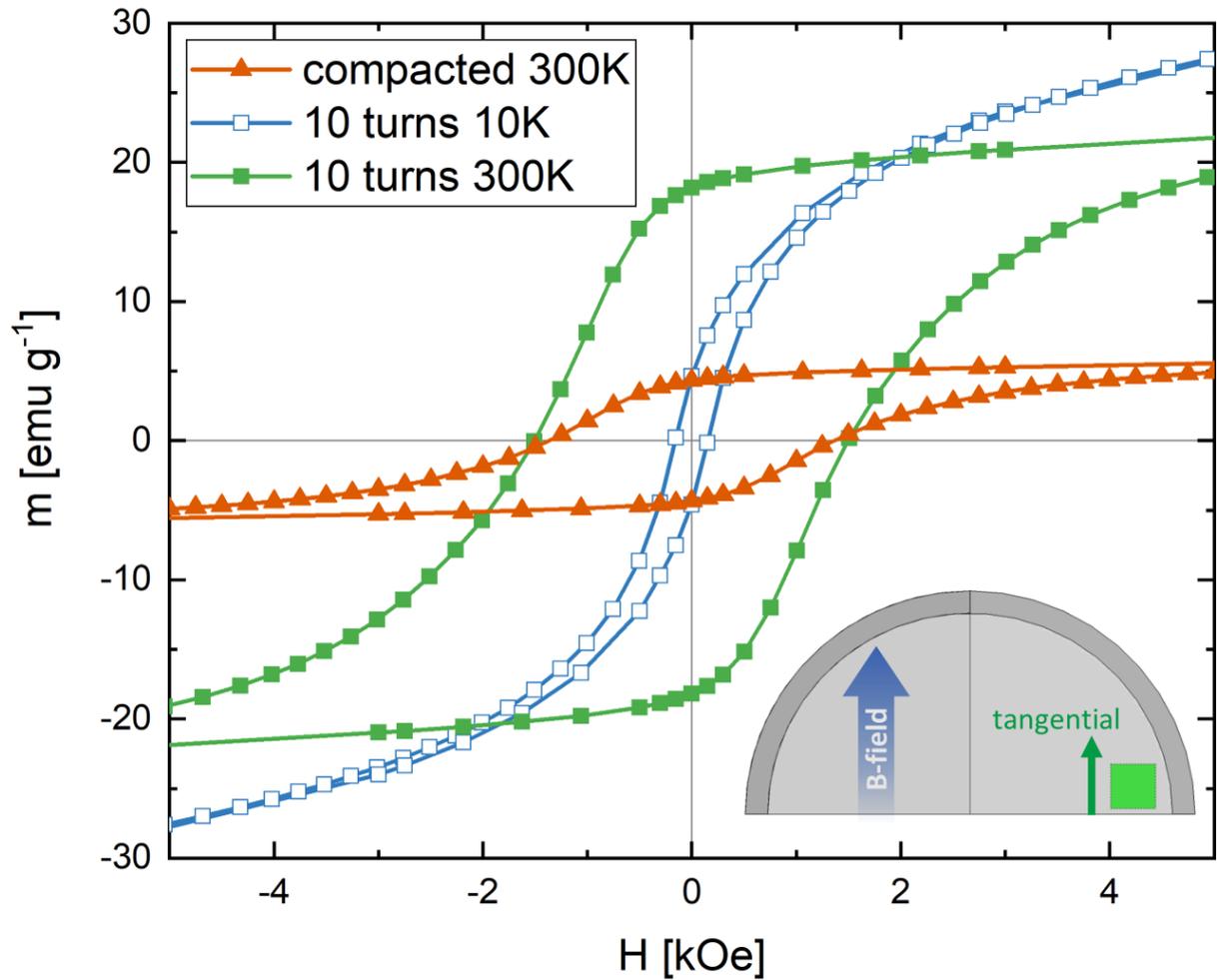

*Figure 6: SQUID magnetometry hysteresis loops of magnetic field assisted vacuum annealed HPT-compacted and HPT-deformed samples. Comparing both processing routes (compaction vs. deformation) $M_s$ substantially rised for the HPT-deformed sample. $H_c$ is reduced for the HPT-deformed sample if measured at 10 K, verifying the existing α-MnBi phase.*

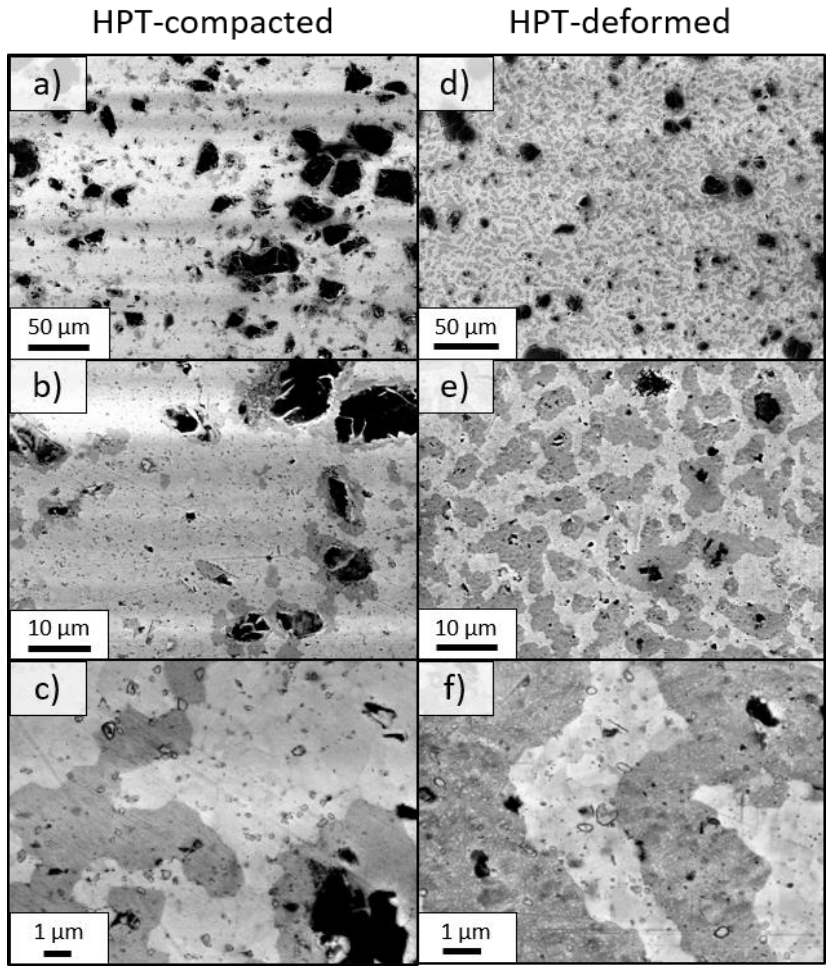

*Figure 7: BSE images with different magnifications of an HPT-compacted sample after MVA a) – c) and an HPT-deformed (10 revolutions) sample after MVA d) – f).*

BSE images with different magnifications of the HPT-compacted and HPT-deformed samples after MVA are shown in Figure 7. The images are recorded in tangential viewing direction with respect to the HPT-disc. It can be seen, that the α-MnBi phase (medium atomic number (Z), medium grey regions) preferentially forms around existing Mn particles (low Z, dark regions). Larger Mn particles and a lower α-MnBi phase fraction is recognized for the HPT-compacted sample. The Bi phase (high Z, bright regions) of the deformed sample still exhibits grain sizes below 1 µm (see Figure 7f) although the composite material is annealed for 4 h at a homologous temperature of $T_{h,Bi}$ = 0.89.

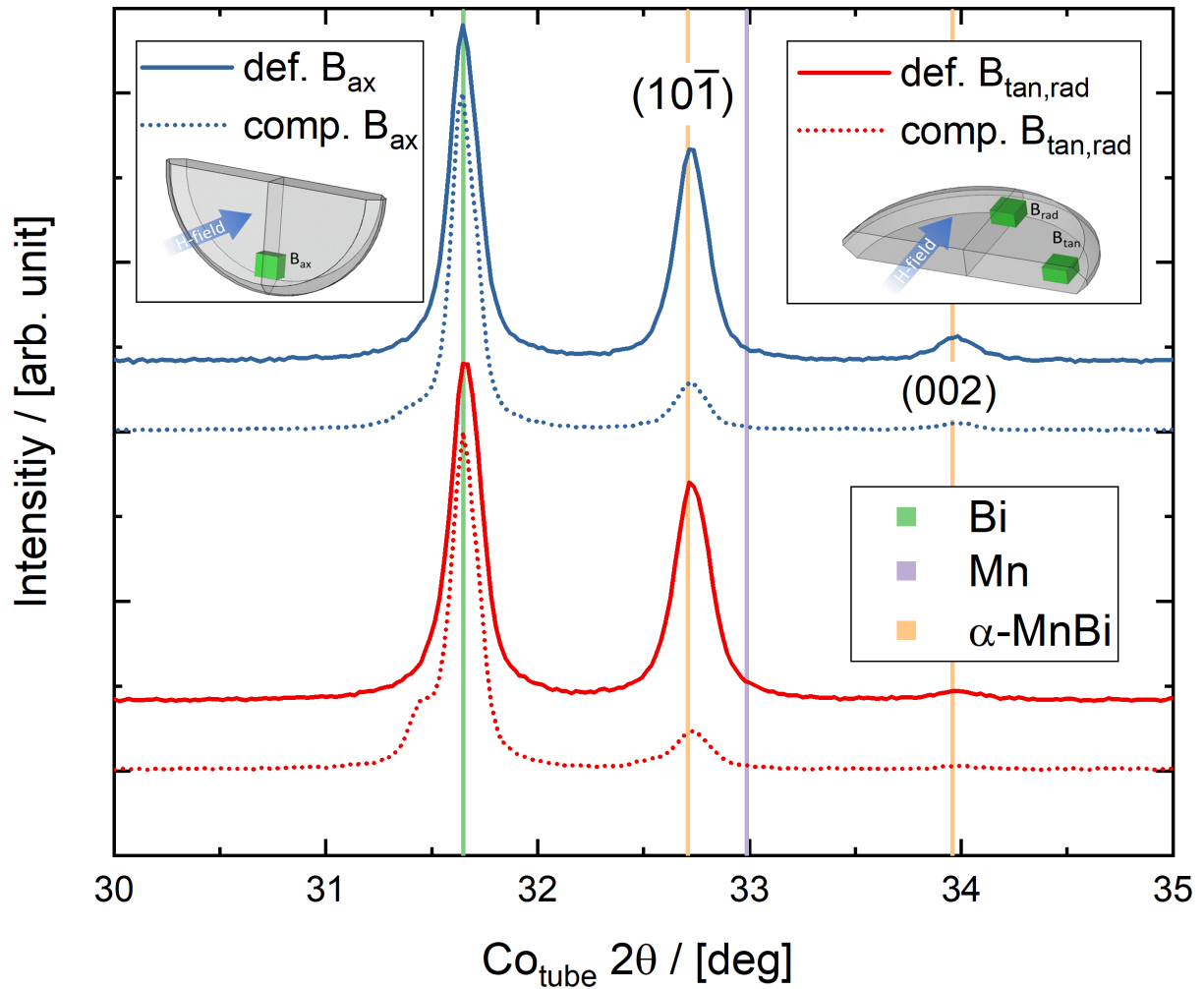

*Figure 8: X-ray diffraction patterns of HPT-compacted (dotted lines) and HPT-deformed (continuous lines) after magnetic field assisted vacuum annealing. The magnetic field is applied perpendicular (blue lines, $B_{ax}$) and parallel (red lines, $B_{tan,rad}$) to the shear deformation. A tiny peak slightly lower to the Bi peak is assigned to a forming oxide, but only for the compacted sample.*

The increased occurrence of the α-MnBi phase formation after HPT-deformation and subsequent MVA is further supported by XRD measurements (Figure 8). The halved HPT-discs are MVA-treated with a field direction applied perpendicular (blue lines, $B_{ax}$) and parallel (red lines, $B_{tan,rad}$) to the HPT-shear deformation. The curves are normalized with respect to the most intense Bi peak at 2θ = 31.65°. Dotted lines represent the HPT-compacted and continuous lines the HPT-deformed sample. In general, pronounced α-MnBi peaks are visible for HPT-deformed sample, whereas the compacted sample exhibits weaker intensities.

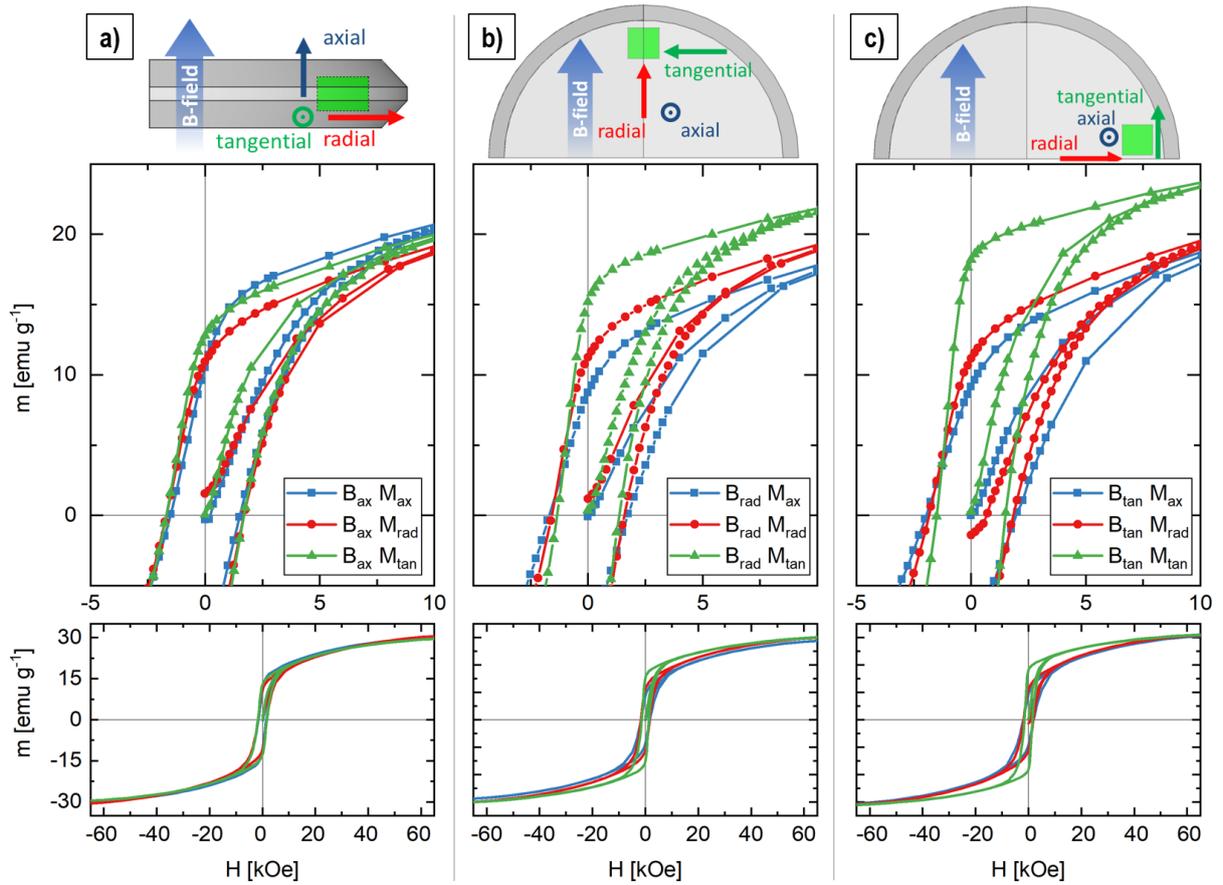

*Figure 9: Hysteresis loops of an HPT-deformed sample after MVA treatment. Cut pieces (green cubes in scheme) undergo the MVA with field direction applied parallel to the a) axial b) radial and c) tangential HPT-disc direction. Sections as well as the complete hysteresis loops are presented.*

These XRD measurements show a strong improvement of α-MnBi phase formation after HPT-deformation and MVA. Thus, magnetic measurements are conducted to an HPT-deformed sample (10 revolutions, 2GPa) after MVA for 4h. The induced texture, which is introduced by HPT, and the direction of MVA are set in correlation with the magnetic properties. To do so, three samples ($B_{ax}$, $B_{rad}$ and $B_{tan}$, compare Figure 1b) are measured by SQUID magnetometry. After measuring one hysteresis loop, the sample is demagnetized thus, the remanence $M_r$ is minimized before the applied measurement field direction is rotated for the next measurement. All principal directions with respect to the HPT-disc (axial, radial, tangential) were aligned with the magnetic field and they are denoted as $M_{ax}$, $M_{rad}$ and $M_{tan}$. This is schematically indicated in Figure 9, where the MVA field direction is shown as a blue arrow and the samples are highlighted (green cubes) within the HPT-disc. The resulting nine hysteresis loops are shown in detail as well as complete loops in Figure 9. The corresponding magnetic properties are summarized in Table 1.

Table 1: Magnetic properties of hysteresis loops presented in Figure 9. Samples with different MVA directions are measured in three perpendicular measurement field directions. The corresponding coercivity $H_c$, saturation magnetization $M_s$, remnant magnetization $M_r$ and the squareness ratio $M_r/M_s$ are listed.

| MVA direction | SQUID direction | $H_c$ [kOe] | $M_s$ [emu g$^{-1}$] | $M_r$ [emu g$^{-1}$] | $M_r/M_s$ |
|---|---|---|---|---|---|
| $B_{ax}$ | $M_{ax}$ | 1.52 | 33.5 | 10.5 | 0.31 |
| $B_{ax}$ | $M_{rad}$ | 1.69 | 35.0 | 10.9 | 0.31 |
| $B_{ax}$ | $M_{tan}$ | 1.69 | 33.1 | 12.7 | 0.38 |
| $B_{rad}$ | $M_{ax}$ | 1.69 | 33.0 | 8.7 | 0.26 |
| $B_{rad}$ | $M_{rad}$ | 1.59 | 34.3 | 11.2 | 0.33 |
| $B_{rad}$ | $M_{tan}$ | 1.37 | 33.4 | 15.1 | 0.45 |
| $B_{tan}$ | $M_{ax}$ | 1.96 | 35.2 | 9.1 | 0.26 |
| $B_{tan}$ | $M_{rad}$ | 1.84 | 35.1 | 11.1 | 0.32 |
| $B_{tan}$ | $M_{tan}$ | 1.52 | 33.9 | 18.1 | 0.53 |

**Discussion**

*Influence of pressure and HPT-deformation on α-MnBi formation*

Similarities between HPT-deformation and ball milling are known. In particular, if applied to a single-phase metallic material, the microstructure is generally refined. Although it is possible to induce very high amounts of strain by extensive milling times or by HPT-deformation, the refinement and resulting grain sizes are limited as a steady state regime is reached [27,35–37]. By combining both techniques, it is possible to add beneficial advantages, e.g., ball milling allows to form a homogeneously powder blend of refined Mn and Bi particles and subsequent HPT-deformation leads to a completely dense bulk material. Therefore, highly uniform samples are processed enabling to profoundly study influences of HPT processing on the α-MnBi formation.

The milling process is often performed with a surfactant meant to further decrease the particle size. However, in a recent study milling of arc melted ingots within an Ar-gas atmosphere was found to be beneficial for the α-MnBi phase [13,38–40]. Therefore, the milling process was executed in Ar-atmosphere. The obtained XRD spectra (Figure 2) show peak broadening at the beginning of the milling process but do not show any significant changes after 2 h. This is explained by the difference in volumetric content of Mn (17 vol.%) and Bi (83 vol.%), to obtain the intended equal atomic ratio. The volumetric excess of the ductile Bi is conceived to cover the Mn particles and hinders their further refinement while the kinetic energy is absorbed by the higher plasticity of the Bi phase. A similar behavior upon ball milling of ductile - brittle material systems, is already reported [41–43]. BSE images (not shown) of the 4 h milled powder blend reveal a homogeneous dispersion of Mn particles, which are reduced from 44 µm to sizes below 1 µm, enclosed by a Bi matrix. The major part of Mn particle refinement is expected to occur during the first 2 h of milling, yet a homogeneous distribution of Mn particles and sizes within a Bi matrix is assured by all means after 4 h of milling.

An influence of the applied pressure during HPT-compaction is found for the α-MnBi phase formation. A faster phase formation is found for the 2 GPa sample, as already after 4 h annealing contrast changes in cigar-shaped regions are found, see Figure 3. After 24 h annealing time, for both samples the α-MnBi phase evolved. This is confirmed by EDX measurements (see Figure 3) and hardness measurements. The measured hardness values for the α-MnBi phase (66 ± 3 HV) are lower as previously reported 109 ± 15 HV [44], however, the microstructure is not comparable. A hardness value similar to the literature is expected when the ratio of α-MnBi phase increases or nanoindentation would have been performed directly at the existing α-MnBi phase. Interestingly, the α-MnBi phase fraction is not increased with further increased annealing times, rather it vanishes. Again, this process is faster for the 2 GPa sample but also the development of an oxide layer on its surface is found (see Supplementary). The preferred formation of an oxide can be explained by the enthalpy of formation, which shows a much lower value for Mn-oxides (and therefore a more stable bond) compared to the α-MnBi phase [45,46]. Furthermore, the dissolution of already existing α-MnBi phases could be driven by an increased Mn diffusion to the sample surface. At least for thin films, a high migration of Mn atoms in Bi is already reported [47]. On the other hand, in case of O diffusion the pronounced oxygen sensitivity of the α-MnBi phase is already described by Cui et al. [48]. However, for the sample compacted at 5 GPa, a similar but decelerated behavior is observed. When applying 5 GPa, the pressure-temperature diagram for pure Bi shows at least two phase transitions [49], which might also occur during the loading and unloading pressure within the compaction process in the current study. As the only difference between the samples is the compaction pressure, probably these phase transitions influence the diffusion of Mn or O in the Bi matrix, even though the pressure is not applied during annealing. A profound understanding of the mentioned diffusion processes requires further studies. In any case, an exposure to oxygen was avoided for all further experiments in this work.

The HEXRD results (Figure 4) confirm, that the α-MnBi phase (represented by the {101} α-MnBi peak, the most intensive peak for polycrystalline powder) is formed for all investigated samples but again at different formation velocities. Verifying the previous results from ex-situ annealing treatments, an accelerated phase formation is found for the 2 GPa HPT-compacted sample compared to the 5 GPa one, requiring about 2000 s and 5000 s to form, respectively. This behavior is even improved further, when HPT-deformed samples are annealed, requiring about 200 s for the {101} α-MnBi peak to form. This can be explained by enhanced diffusion processes, as already known for materials obtained by severe plastic deformation [50,51]. We will discuss this topic in more detail later on. It is more difficult to understand the relative intensities within the presented figures. Based on the crystal structure (COD: 9011108 [30]), the intensity of the {220} Mn-peak is about 2.4 % of the most pronounced Mn-peak ({411} at $q = 29.9$ nm$^{-1}$) in an isotropic polycrystalline sample. Although being present in all spectra, the {220} signal is low. Yet due to the normalization to the maximum peak value ({002} α-MnBi for 5GPa HPT-compacted and {101} α-MnBi else) of every given image in Figure 4, it appears with a different intensity. This is, because the {101} α-MnBi peak grows to a different extend during annealing and this also explains the seemingly increased background signal for both HPT-deformed samples. However, at a first sight, an analysis of the diffraction pattern did not show a clear correlation between peak intensities and processing routes.

It appears, that the integrated intensities and thus supposed amount of α-MnBi phase strongly depends on the phase formation characteristics. When we compare the 2D-detector recordings of the 3 h state with the as-processed state in Figure 5, bright spots evolve midst the beforehand existing Mn and Bi rings for both compacted samples. At exactly the same diffraction angles where bright spots are visible, continuous lines are found for the HPT-deformed samples. Consequently, the α-MnBi phase forms in larger crystallites within the HPT-compacted samples (few, larger grains), whereas the HPT-deformed samples show a more homogeneously distributed nucleation behavior, which seems to be independent of the pressure applied. When the samples are annealed for 24 h and measured ex-situ by HEXRD, as shown in the third row of Figure 5, no changes regarding the number of nucleation sites are detected as indicated by a similar number of α-MnBi spots in the diffraction patterns. Besides the conclusions on the differences in nucleation behavior, the different appearance of α-MnBi in the diffraction patterns of HPT-compacted and HPT-deformed samples also has consequences on the interpretability of the peak intensities. Since the integrated intensities for the HPT-compacted samples originate from very few high-intensity spots of large α-MnBi crystallites and the presence or absence of a single crystallite with measurable orientation has a huge effect, these intensities are not used for conclusions about phase fractions due to the enormous statistical error. This is also illustrated by the fact that the {002} α-MnBi peak for the 5 GPa compacted sample is higher than in any other measurement, but is based on a single crystallite with accidentally beneficial orientation.

The statistics of these experiments is given by the diffraction volume and the size of the crystallites. The samples are measured in, with a minimum thickness of about 500 µm for HPT-deformed samples and about 800 µm for compacted samples, at a spot size of 200x200 µm$^2$. Comparing to the 4h and 24 h images in Figure 3, several α-MnBi phase regions are expected to exist in this interaction volume. Moreover, when the annealing times are increased, the number of detected α-MnBi phase spots is barely increased, showing that grain growth is preferred over nucleation at this stage. It is also possible to distinguish between the different compaction pressures, as for the 5 GPa compacted sample a higher number of α-MnBi phase spots is visible similar to the higher number of "cigar-shaped" regions in SEM (compare to Figure 3). HPT-deformation seems to be beneficial as the continuous Debye-Scherrer rings indicate a considerable higher density of α-MnBi phase nucleation sites. Diffusion mechanism are the key aspect for the α-MnBi phase to form. The annealing temperature in the experiments is constantly kept below the eutectic temperature of 265 °C [52,53], which is relatively low for diffusion processes. However, the high defect density induced by plastic deformation could enhance the inter-diffusion process of Mn and Bi, as such behavior is already reported for Ni [54–56]. Additionally, this phase forms almost directly after the heat treatment starts further sustaining this argument.

Consequently, the most important findings from these measurements are as follwed:
- A combination of ball milling, HPT and annealing leads to α-MnBi phase formation.
- HPT-deformation prior annealing treatments drastically enhances the number of α-MnBi phase nucleation sites.
- Diffusion processes as most decisive phase formation parameter is correlated to the defect density
- Long time annealing at ambient conditions adversely affects the α-MnBi phase formation.
- α-MnBi grain growth is preferred over nucleation, at least for HPT-compacted samples.
- HPT-processing at 2 GPa yields a higher α-MnBi phase fraction than at 5 GPa.

*Influence of magnetic field assisted annealing on the α-MnBi phase formation*

In the second part of this study, the α-MnBi phase formation is further studied. In this part, the high defect density is solely applied by HPT without ball milling and the heat treatment is combined with an additional magnetic field. As stated in the introduction, a commercial use of MnBi based permanent magnets is mainly barred by the inability to manufacture large quantities of bulk α-MnBi phase. In addition, porosity often is a limiting factor for magnetic properties when starting with powders. However, HPT-processing of metallic powders lead to completely dense materials. The measured density for the HPT-compacted (8.778 ± 0.223 g/cm$^3$) is lower compared to the HPT-deformed (9.421 ± 0.221 g/cm$^3$) sample. The density value of the compacted powder sample is in good agreement with the calculated value of 8.625 g/cm$^3$ using theoretical values and a medium composition. The higher density for the HPT-deformed sample could indicate a preferred Mn particle transport towards the HPT-disc edge during deformation, thus leading to a higher Bi content. However, a significant porosity of HPT treated samples is excluded. To increase the α-MnBi volume fraction, it is necessary to rise the number of nucleation sites and additionally accelerate phase growth. A beneficial influence of magnetic field annealing on α-MnBi phase formation, was already mentioned in 1958 [57], although its origin is not fully understood until today. It is stated that field annealing increases the Zeeman energy affecting the critical radius of nuclei. Furthermore, a decrease of free energy for the α-MnBi phase formation is reported to result in a negative formation enthalpy and an enhanced nucleation rate [22,23]. The resulting localized reaction heat is believed to be responsible that Mn and Bi particles partially melt. Thus, a rotation of crystallites occurs leading to a textured and anisotropic phase formation if a magnetic field is applied [15,58,59]. Moreover, the huge magnetocrystalline anisotropy of about 2.2 MJ/m$^3$ at the annealing temperature (herein 240 °C) seems to affect the recrystallisation process (including nucleation and grain growth) [6,10] and analogies to strain-induced recrystallisation can be drawn [59,60]. In general, magnetic field annealing of a Mn-Bi composite leads to an enhanced α-MnBi phase volume fraction and grain alignment (texture formation) [15,23,58,61], which is also supported by the results of the current study.

The volume fraction of α-MnBi can be calculated with the measured saturation magnetization $M_s$ as it is proportional to the α-MnBi fraction, considering the theoretical $M_s$ of 79 emu/g [9,12]. Therefore, we can estimate the volume content by using the calculated $M_s$ value of Figure 6 and the theoretical value [12]. The amount of α-MnBi phase in the HPT-compacted sample after MVA equals to a content of 12 %. Only by deforming the sample prior to MVA, the volume fraction increases above 50 % ($M_s$ = 33.92 emu/g) indicating a higher number of nucleation sites. Except for the applied strain, no differences between HPT-compacted and HPT-deformed samples shown in Figure 6 are present. The higher strain leads to a higher number of small Mn particles, increasing the size of Mn and Bi interfaces. Furthermore, the microstructural defect density ρ is increased, which is related to the critical nucleation radius $r_c = \frac{2\gamma}{p}$. γ is the specific grain boundary energy and p is the driving force, whereby p is directly proportional to ρ [62]. Therefore, with $r_c$ is lowered with increasing ρ. Consequently, a larger phase interface and high ρ – both tunable by HPT – enhances the α-MnBi phase formation. This is supported by the measured hysteresis loops and by BSE images presented in Figure 7.

The $H_c$ of the hysteresis loops (Figure 6) is strongly decreased when measured at lower temperatures, another strong indication for the presence of the α-MnBi phase due to its positive temperature coefficient of intrinsic coercivity [8,9]. The $H_c$ measured at 300 K of the HPT-compacted and HPT-deformed sample are comparable: 1.38 kOe and 1.58 kOe, respectively. However, lower values can be found in the literature after similar field annealing measurements. For example, Gabay et al. [15] found a $H_c$ of 0.6 kOe. Compressed MnBi powder blends in a study of Mitsui et al. [58] attain a $H_c$ of 0.8 kOe. Thus, HPT-deformation has a beneficial effect on magnetic properties. Magnetic hardening by grain refinement is well known and it is thought to mainly contribute to a change of $H_c$ of the α-MnBi phase particles [63–65]. This is imposingly demonstrated by Mitsui et al. [22], where magnetic field annealing leads not only to an enhanced volume of α-MnBi phase but also to a decreased $H_c$ due to an increased grain size after annealing. Based on the phenomenologically described expression $H_c = H_a \left(\frac{\delta}{D}\right)^n$ [65], the α-MnBi phase of the HPT-deformed samples show crystallite sizes D of about 0.7 µm, whereby δ is the Bloch wall width (7 nm), $H_a$ the theoretical limit of $H_c$ at D ≈ δ (40 kOe) and an exponent n (0.7) is used [10,65]. For such crystallite sizes a large $H_c$ (> 10 kOe) could be expected [48,66].

Based on the above introduced expression for $H_c$, two different samples (HPT-compacted and HPT-deformed) only exhibit a similar $H_c$ if the crystallite size is similar. Combining this information with the higher amount of α-MnBi phase fraction present for the HPT-deformed sample, this further supports our results obtained from HEXRD, where HPT-deformation increases the number of α-MnBi nucleation sites. Accordingly, it is noted that magnetic properties are mainly influenced during the initial annealing process (regarding magnetic field annealing procedures) and changes stagnate for increased annealing times [15,22,60,67].

Concluding, primarily effects of α-MnBi phase formation are due to the enhanced nucleation behavior after HPT-deformation and higher amounts of defects. Additionally, the reduced grain size of the Mn-Bi composite after HPT, thus enlarged grain boundary volume and defect density favors grain boundary diffusion and facilitates phase formation [54–56]. The resulting α-MnBi crystallite sizes seem to be small and independent of previous HPT-deformation.

*HPT-induced magnetic anisotropy*

HPT-deformation often induces texture in metallic materials [68]. For hard magnetic $SmCo_5$ composites, which were HPT-deformed, a textured microstructure and as a result, anisotropic magnetic behavior is reported [29,69]. Thus, an influence of differently aligned MVAs in combination with the crystallographic texture of the HPT-deformed MnBi sample is expected. However, the XRD measurements on the $B_{ax}$ and $B_{tan,rad}$ annealed samples (Figure 8) reveal similar texture for the Bi phase after MVA for differently applied field directions, since the intensities of all Bi peaks are similar in the two samples. Due to the low peak intensities of Mn an analysis of this phase is omitted. However, differences are found for the α-MnBi peaks of the HPT-deformed sample. For the $B_{ax}$ anneal the (002) and (004) planes and for the $B_{tan,rad}$ anneal the (110) plane are more pronounced, respectively. This fits to recently reported findings, where the c-axis of the α-MnBi phase tend to form parallel to the applied magnetic field [59,61,70–72].

Based on SQUID measurements, a detailed investigation of magnetic properties is presented in Figure 9 and corresponding magnetic parameters are summarized in Table 1. In general, a higher remnant magnetization $M_r$ indicates an increased amount of crystallographically aligned grains [59,61]. The $M_s$ value slightly differ between 33.0 emu g$^{-1}$ and 35.2 emu g$^{-1}$. Thus, the α-MnBi phase content varies between 41.8 % and 44.6 %. Therefore, the amount of forming α-MnBi phase is not affected by different MVA directions though, magnetic properties, e.g., $H_c$ and $M_r$ are influenced. Further commenting on the differences of the α-MnBi phase content, which can occur, if α-MnBi phase particles of sizes in the range of a quasi-single domain regime, exhibiting a high $H_c$, are not completely demagnetized between the presented $M_{ax}$, $M_{tan}$ and $M_{rad}$ measurements. Therefore, the sample exhibits residual magnetic fields, when it is rotated. This could be circumvented, if magnetic measurement fields would be applied at higher magnetic fields, which is not possible using the current device.

For the $B_{ax}$ sample (Figure 9a), the measured values for $H_c$ and $M_r$ show an isotropic magnetic behavior of the α-MnBi phase. In contrast, $H_c$ and $M_r$ change for the $B_{rad}$ sample (Figure 9b) as a function of the SQUID measuring field direction. Similar behavior is observed for the $B_{tan}$ sample (Figure 9c), which show the largest differences. If the α-MnBi phase formation would be only dependent on the direction of the applied magnetic field during MVA, one would expect hysteresis loops where the easy and hard axis of the α-MnBi phase can be correlated to the MVA field. This is already reported and explained due to a partially melting of the surrounding Bi and an enabled rotation of the α-MnBi grains [15,23,58,61]. However, this is clearly not the case for our samples as the $M_r$ value does not follow the applied field directions. Thus, a significant influence of the applied shear deformation during HPT-deformation is presumed. The squareness ratio $M_r/M_s$ (presented in Table 1) is a frequently used measure for degree of texture, thus the anisotropic magnetic behavior. It is highest when the measuring field direction is applied in tangential HPT-disc direction (anisotropic) and lowest for the axial direction (isotropic), a valid observation for all field annealing directions. This observation is assumed to originate in the shear deformed microstructure, entailing a preferred orientation for the forming α-MnBi phase. For thin films, the α-MnBi grains are reported to preferably align their c-axis parallel to the Bi c-axis [67,73]. As the Bi phase is the one that is easily accessible by microstructural investigations and to check if the α-MnBi phase formation correlates with a preexisting texture of the MnBi composite prior annealing, a pure Bi sample is processed by HPT-deformation (RT, 25 rotations, 2 GPa). The performed EBSD scans are recorded in tangential HPT-disc viewing direction. Exemplarily an inverse pole figure map is presented in Figure 10a) and a pole figure obtained from several scans is shown in Figure 10b). The radial and tangential HPT-disc directions point in horizontal and vertical direction, while the axial HPT-disc direction points perpendicular to the paper plane. A largely identical orientation of the Bi grains is visible, with the c-axis of the Bi crystals pointing parallel to the axial HPT-disc direction. Assuming the same texture in the composite sample, the α-MnBi grains should be perfectly aligned when the MVA field is applied parallel to the crystallographic textured Bi, thus the axial HPT-disc direction. Yet, such a behavior is not observed by our measurements for the $B_{ax}$ sample.

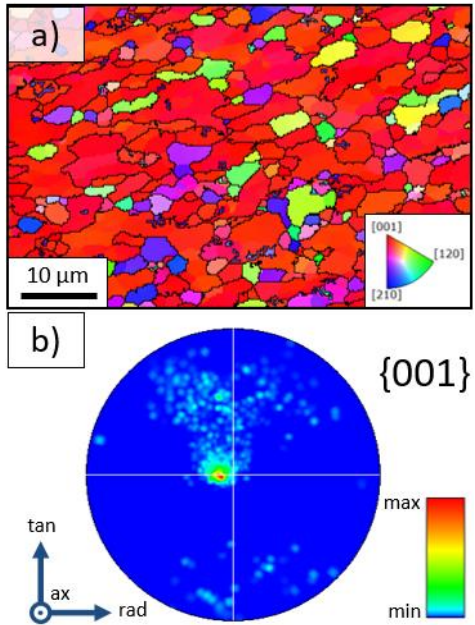

*Figure 10: a) Inverse pole figure map and b) corresponding pole figure for the {001} plane obtained from EBSD scans on HPT-deformed pure Bi sample (RT, 25 rotations, 2 GPa). For the pole figure, the radial and tangential HPT-disc directions point in horizontal and vertical direction. The axial direction points perpendicular to the paper plane. The maximum intensity in the pole figure correlates to the axial HPT-disc direction.*

This clearly shows the importance of diffusion processes on α-MnBi formation enabled by HPT-deformation. The initial phase nucleation is assumed to happen at Mn and Bi particle interfaces [23]. Diffusion processes in solids prefer to occur along grain boundaries rather through the interiors of crystals. An enhanced grain boundary diffusion is known for HPT materials [54–56] and found for polycrystalline Bi [74], as well as Mn diffusion in Bi thin films [47]. Moreover, in slightly different experiments, we obtained similar results as Borsup et al. [66] where a rapid and incredibly extensive diffusion of molten Bi in polycrystalline Mn is found (see Figure Supplementary 2).

The highest magnetic anisotropy is found for $B_{tan}$. If grain boundaries are aligned parallel to the MVA field (compare to Figure 10a), the forming phase could preferably grow along this direction (as mass transport is relieved) and simultaneously form textured grains. Additionally, one should think about the textured Bi-matrix. The surface normal of the preferred (001) slip plane [75] is parallel to the c-axis, allowing dislocation pathways within the shear-plane. Therefore, so called short-circuit or pipe diffusion [76] is directed within the shear plane, additionally enhancing diffusion processes in the mentioned plane. The formation of a textured α-MnBi phase upon MVA seems to be facilitated by diffusion processes in the shear-plane, whereas α-MnBi crystallite growth upon MVA in axial HPT-disc direction thus, normal to the dislocation pathways and grain boundaries, is suppressed. Thereby, the direction of applied MVA can be used to tune the grade of anisotropy of hard magnetic properties within the HPT-processed sample.

**Conclusion**

In this work, we successfully produced the ferromagnetic α-MnBi phase for the first time by applying severe plastic deformation using HPT and a subsequent thermal processing step. The influence of applied pressure by HPT changes the rate of the α-MnBi phase formation, as well as its decomposition if annealed at atmosphere. The initial phase nucleation and formation is studied by means of in-situ HEXRD annealing and it is found that the α-MnBi phase forms shortly after beginning of the heat treatment for HPT-deformed samples. This process depends on the applied pressure and is slower for HPT-compacted samples without severe deformation, thus lower microstructural defects. Most importantly, a correlation between higher α-MnBi phase nucleation sites and applied HPT-deformation is found, further revealing that upon annealing α-MnBi phase growth is preferred over nucleation at least for the HPT-compacted samples. By applying magnetic field assisted vacuum annealing (MVA) to HPT-processed samples, the α-MnBi phase formation is enhanced. HPT-compacted and HPT-deformed samples show similar coercivities, thus similar α-MnBi grain sizes, but a differing saturation magnetization as the α-MnBi phase reaches a higher volume fraction (> 50 vol%) for the HPT-deformed one. Our results point towards the importance of diffusion processes for the α-MnBi phase formation. The diffusion is enhanced for samples with higher defect densities (HPT-deformed compared to HPT-compacted) and influenced by the shear deformation induced, refined grain boundary structure. This further provides the possibility to form anisotropic hard magnets. Summarized, the α-MnBi phase formation is best for HPT-deformed and MVA samples, where the applied annealing field is parallel to the HPT shear direction.

**Declaration of Competing Interest.** The authors declare, that they have no known competing financial interests or personal relationships, that could have appeared to influence the work reported in this paper.

**Acknowledgement.** We acknowledge DESY (Hamburg, Germany), a member of the Helmholtz Association HGF, for the provision of experimental facilities. Parts of this research were carried out at beamline P21.2 at PETRA III under proposal I-20190577.

This project has received funding from the European Research Council (ERC) under the EuropeanUnion's Horizon 2020 research and innovation programme (Grant No. 757333 and 101069203).

**References**

[1] N. Makino, A Review of Studies on the Permanent Magnet, Tetsu-to-Hagane. 44 (1958) 1217–1224. https://doi.org/10.2355/tetsutohagane1955.44.10_1217.
[2] E.W. Lee, L. F. Bates, Nature. 272 (1978) 568–568. https://doi.org/10.1038/272568a0.
[3] T. Chen, Contribution to the equilibrium phase diagram of the Mn–Bi system near MnBi, Journal of Applied Physics. 45 (1974) 2358–2360. https://doi.org/10.1063/1.1663594.
[4] K. Oikawa, Y. Mitsui, K. Koyama, K. Anzai, Thermodynamic Assessment of the Bi-Mn System, Mater. Trans. 52 (2011) 2032–2039. https://doi.org/10.2320/matertrans.M2011229.
[5] Y. Mitsui, K. Oikawa, K. Koyama, K. Watanabe, Thermodynamic assessment for the Bi–Mn binary phase diagram in high magnetic fields, Journal of Alloys and Compounds. 577 (2013) 315–319. https://doi.org/10.1016/j.jallcom.2013.05.198.
[6] J. Cui, M. Kramer, L. Zhou, F. Liu, A. Gabay, G. Hadjipanayis, B. Balasubramanian, D. Sellmyer, Current progress and future challenges in rare-earth-free permanent magnets, Acta Materialia. 158 (2018) 118–137. https://doi.org/10.1016/j.actamat.2018.07.049.
[7] D.T. Zhang, W.T. Geng, M. Yue, W.Q. Liu, J.X. Zhang, J.A. Sundararajan, Y. Qiang, Crystal structure and magnetic properties of $Mn_xBi_{100-x}$ (x=48, 50, 55 and 60) compounds, Journal of Magnetism and Magnetic Materials. 324 (2012) 1887–1890. https://doi.org/10.1016/j.jmmm.2012.01.017.
[8] J. Cui, J.P. Choi, G. Li, E. Polikarpov, J. Darsell, N. Overman, M. Olszta, D. Schreiber, M. Bowden, T. Droubay, M.J. Kramer, N.A. Zarkevich, L.L. Wang, D.D. Johnson, M. Marinescu, I. Takeuchi, Q.Z. Huang, H. Wu, H. Reeve, N.V. Vuong, J.P. Liu, Thermal stability of MnBi magnetic materials, J. Phys.: Condens. Matter. 26 (2014) 064212. https://doi.org/10.1088/0953-8984/26/6/064212.
[9] J.B. Yang, Y.B. Yang, X.G. Chen, X.B. Ma, J.Z. Han, Y.C. Yang, S. Guo, A.R. Yan, Q.Z. Huang, M.M. Wu, D.F. Chen, Anisotropic nanocrystalline MnBi with high coercivity at high temperature, Appl. Phys. Lett. 99 (2011) 082505. https://doi.org/10.1063/1.3630001.
[10] X. Guo, X. Chen, Z. Altounian, J.O. Ström-Olsen, Magnetic properties of MnBi prepared by rapid solidification, Phys. Rev. B. 46 (1992) 14578–14582. https://doi.org/10.1103/PhysRevB.46.14578.
[11] T. Chen, W. Stutius, The phase transformation and physical properties of the MnBi and $Mn_{1.08}Bi$ compounds, IEEE Transactions on Magnetics. 10 (1974) 581–586. https://doi.org/10.1109/TMAG.1974.1058367.

[12] J. Park, Y.-K. Hong, J. Lee, W. Lee, S.-G. Kim, C.-J. Choi, Electronic Structure and Maximum Energy Product of MnBi, Metals. 4 (2014) 455–464. https://doi.org/10.3390/met4030455.

[13] N.M. Lam, T.M. Thi, P.T. Thanh, N.H. Yen, N.H. Dan, Fabrication of Mn-Bi Nanoparticles by High Energy Ball Milling, Mater. Trans. 56 (2015) 1394–1398. https://doi.org/10.2320/matertrans.MA201577.

[14] N.V. Rama Rao, A.M. Gabay, X. Hu, G.C. Hadjipanayis, Fabrication of anisotropic MnBi nanoparticles by mechanochemical process, Journal of Alloys and Compounds. 586 (2014) 349–352. https://doi.org/10.1016/j.jallcom.2013.10.067.

[15] A.M. Gabay, G.C. Hadjipanayis, J. Cui, New anisotropic MnBi permanent magnets by field-annealing of compacted melt-spun alloys modified with Mg and Sb, Journal of Magnetism and Magnetic Materials. 495 (2020) 165860. https://doi.org/10.1016/j.jmmm.2019.165860.

[16] A.M. Gabay, G.C. Hadjipanayis, J. Cui, Effect of Sb substitution on crystal structure, texture and hard magnetic properties of melt-spun MnBi alloys, Journal of Alloys and Compounds. 792 (2019) 77–86. https://doi.org/10.1016/j.jallcom.2019.03.407.

[17] Y.B. Yang, X.G. Chen, S. Guo, A.R. Yan, Q.Z. Huang, M.M. Wu, D.F. Chen, Y.C. Yang, J.B. Yang, Temperature dependences of structure and coercivity for melt-spun MnBi compound, Journal of Magnetism and Magnetic Materials. 330 (2013) 106–110. https://doi.org/10.1016/j.jmmm.2012.10.046.

[18] W. Tang, G. Ouyang, X. Liu, J. Wang, B. Cui, J. Cui, Engineering microstructure to improve coercivity of bulk MnBi magnet, Journal of Magnetism and Magnetic Materials. 563 (2022) 169912. https://doi.org/10.1016/j.jmmm.2022.169912.

[19] P. Kainzbauer, K.W. Richter, H.S. Effenberger, G. Giester, H. Ipser, The Ternary Bi-Mn-Sb Phase Diagram and the Crystal Structure of the Ternary T Phase Bi0.8MnSb0.2, J. Phase Equilib. Diffus. 40 (2019) 462–481. https://doi.org/10.1007/s11669-019-00719-x.

[20] S. Lu, S. Shuai, L. Chen, Z. Xiang, W. Lu, Effect of Mg content on the microstructure and magnetic properties of rare-earth-free MnBi alloys, Journal of Magnetism and Magnetic Materials. 570 (2023) 170499. https://doi.org/10.1016/j.jmmm.2023.170499.

[21] H. Kronmüller, J.B. Yang, D. Goll, Micromagnetic analysis of the hardening mechanisms of nanocrystalline MnBi and nanopatterned FePt intermetallic compounds, J. Phys.: Condens. Matter. 26 (2014) 064210. https://doi.org/10.1088/0953-8984/26/6/064210.

[22] Y. Mitsui, R.Y. Umetsu, K. Takahashi, K. Koyama, Reactive sintering process of ferromagnetic MnBi under high magnetic fields, Journal of Magnetism and Magnetic Materials. 453 (2018) 231–235. https://doi.org/10.1016/j.jmmm.2018.01.026.

[23] D. Miyazaki, Y. Mitsui, R.Y. Umetsu, K. Takahashi, S. Uda, K. Koyama, Enhancement of the Phase Formation Rate during In-Field Solid-Phase Reactive Sintering of Mn-Bi, Mater. Trans. 58 (2017) 720–723. https://doi.org/10.2320/matertrans.MBW201609.

[24] Z. Horita, Y. Tang, T. Masuda, Y. Takizawa, Severe Plastic Deformation under High Pressure: Upsizing Sample Dimensions, Mater. Trans. 61 (2020) 1177–1190. https://doi.org/10.2320/matertrans.MT-M2020074.

[25] H.P. Stüwe, Equivalent Strains in Severe Plastic Deformation, Adv. Eng. Mater. 5 (2003) 291–295. https://doi.org/10.1002/adem.200310085.

[26] K.S. Kormout, R. Pippan, A. Bachmaier, Deformation-Induced Supersaturation in Immiscible Material Systems during High-Pressure Torsion: Deformation-Induced Supersaturation, Adv. Eng. Mater. 19 (2017) 1600675. https://doi.org/10.1002/adem.201600675.

[27] A. Hohenwarter, A. Bachmaier, B. Gludovatz, S. Scheriau, R. Pippan, Technical parameters affecting grain refinement by high pressure torsion, International Journal of Materials Research. 100 (2009) 1653–1661. https://doi.org/10.3139/146.110224.

[28] G. Faraji, H. Torabzadeh, An Overview on the Continuous Severe Plastic Deformation Methods, Mater. Trans. 60 (2019) 1316–1330. https://doi.org/10.2320/matertrans.MF201905.

[29] L. Weissitsch, M. Stückler, S. Wurster, J. Todt, P. Knoll, H. Krenn, R. Pippan, A. Bachmaier, Manufacturing of Textured Bulk Fe-SmCo5 Magnets by Severe Plastic Deformation, Nanomaterials. 12 (2022) 963. https://doi.org/10.3390/nano12060963.

[30] S. Gražulis, A. Daškevič, A. Merkys, D. Chateigner, L. Lutterotti, M. Quirós, N.R. Serebryanaya, P. Moeck, R.T. Downs, A. Le Bail, Crystallography Open Database (COD): an open-access collection of crystal structures and platform for world-wide collaboration, Nucleic Acids Research. 40 (2012) D420–D427. https://doi.org/10.1093/nar/gkr900.

[31] Z. Hegedüs, T. Müller, J. Hektor, E. Larsson, T. Bäcker, S. Haas, A. Conceiçao, S. Gutschmidt, U. Lienert, Imaging modalities at the Swedish Materials Science beamline at PETRA III, IOP Conf. Ser.: Mater. Sci. Eng. 580 (2019) 012032. https://doi.org/10.1088/1757-899X/580/1/012032.

[32] G. Ashiotis, A. Deschildre, Z. Nawaz, J.P. Wright, D. Karkoulis, F.E. Picca, J. Kieffer, The fast azimuthal integration Python library: *pyFAI*, J Appl Crystallogr. 48 (2015) 510–519. https://doi.org/10.1107/S1600576715004306.

[33] C.P. Bean, I.S. Jacobs, Magnetic Granulometry and Super-Paramagnetism, Journal of Applied Physics. 27 (1956) 1448–1452. https://doi.org/10.1063/1.1722287.

[34] D.-X. Chen, A. Sanchez, E. Taboada, A. Roig, N. Sun, H.-C. Gu, Size determination of superparamagnetic nanoparticles from magnetization curve, Journal of Applied Physics. 105 (2009) 083924. https://doi.org/10.1063/1.3117512.

[35] R. Pippan, S. Scheriau, A. Taylor, M. Hafok, A. Hohenwarter, A. Bachmaier, Saturation of Fragmentation During Severe Plastic Deformation, Annu. Rev. Mater. Res. 40 (2010) 319–343. https://doi.org/10.1146/annurev-matsci-070909-104445.

[36] O. Renk, R. Pippan, Saturation of Grain Refinement during Severe Plastic Deformation of Single Phase Materials: Reconsiderations, Current Status and Open Questions, Mater. Trans. 60 (2019) 1270–1282. https://doi.org/10.2320/matertrans.MF201918.

[37] C. Knieke, M. Sommer, W. Peukert, Identifying the apparent and true grinding limit, Powder Technology. 195 (2009) 25–30. https://doi.org/10.1016/j.powtec.2009.05.007.

[38] H.-R. Shen, Z. Yu, Y.-H. Tu, Q. Wu, Z. Wang, Y. Zhao, H.-L. Ge, Microstructure and magnetic properties of anisotropic Mn-Bi powder prepared by low energy ball milling assisted with polyvinylpyrrolidone, Journal of Magnetism and Magnetic Materials. 551 (2022) 169108. https://doi.org/10.1016/j.jmmm.2022.169108.

[39] K. Kanari, C. Sarafidis, M. Gjoka, D. Niarchos, O. Kalogirou, Processing of magnetically anisotropic MnBi particles by surfactant assisted ball milling, Journal of Magnetism and Magnetic Materials. 426 (2017) 691–697. https://doi.org/10.1016/j.jmmm.2016.10.141.

[40] J. Cao, Y.L. Huang, Y.H. Hou, Z.Q. Shi, X.T. Yan, Z.C. Zhong, G.P. Wang, Microstructure and magnetic properties of MnBi alloys with high coercivity and significant anisotropy prepared by surfactant assisted ball milling, Journal of Magnetism and Magnetic Materials. 473 (2019) 505–510. https://doi.org/10.1016/j.jmmm.2018.10.052.

[41] M. Abareshi, S.M. Zebarjad, E.K. Goharshadi, Study of the morphology and granulometry of polyethylene-clay nanocomposite powders, J Vinyl Addit Technol. 16 (2010) 90–97. https://doi.org/10.1002/vnl.20217.

[42] M.S. El-Eskandarany, Mechanical alloying for fabrication of advanced engineering materials, Noyes Publications, Norwich, N.Y, 2001.

[43] R. Taherzadeh Mousavian, R. Azari Khosroshahi, S. Yazdani, D. Brabazon, A.F. Boostani, Fabrication of aluminum matrix composites reinforced with nano- to micrometer-sized SiC particles, Materials & Design. 89 (2016) 58–70. https://doi.org/10.1016/j.matdes.2015.09.130.

[44] X. Jiang, T. Roosendaal, X. Lu, O. Palasyuk, K.W. Dennis, M. Dahl, J.-P. Choi, E. Polikarpov, M. Marinescu, J. Cui, Mechanical and electrical properties of low temperature phase MnBi, Journal of Applied Physics. 119 (2016) 033903. https://doi.org/10.1063/1.4939811.

[45] C.P. Wang, W.J. Yu, Z.S. Li, X.J. Liu, A.T. Tang, F.S. Pan, Thermodynamic assessments of the Bi–U and Bi–Mn systems, Journal of Nuclear Materials. 412 (2011) 66–71. https://doi.org/10.1016/j.jnucmat.2011.02.021.

[46] K.T. Jacob, A. Kumar, G. Rajitha, Y. Waseda, Thermodynamic Data for Mn3O4, Mn2O3 and MnO2, High Temperature Materials and Processes. 30 (2011). https://doi.org/10.1515/htmp.2011.069.

[47] Y. Iwama, U. Mizutani, F. Humphrey, Formation process of MnBi thin films, IEEE Trans. Magn. 8 (1972) 487–489. https://doi.org/10.1109/TMAG.1972.1067370.

[48] J. Cui, J.-P. Choi, E. Polikarpov, M.E. Bowden, W. Xie, G. Li, Z. Nie, N. Zarkevich, M.J. Kramer, D. Johnson, Effect of composition and heat treatment on MnBi magnetic materials, Acta Materialia. 79 (2014) 374–381. https://doi.org/10.1016/j.actamat.2014.07.034.

[49] H.-Y. Chen, S.-K. Xiang, X.-Z. Yan, L.-R. Zheng, Y. Zhang, S.-G. Liu, Y. Bi, Phase transition of solid bismuth under high pressure, Chinese Phys. B. 25 (2016) 108103. https://doi.org/10.1088/1674-1056/25/10/108103.

[50] K. Oh-ishi, K. Edalati, H.S. Kim, K. Hono, Z. Horita, High-pressure torsion for enhanced atomic diffusion and promoting solid-state reactions in the aluminum–copper system, Acta Materialia. 61 (2013) 3482–3489. https://doi.org/10.1016/j.actamat.2013.02.042.

[51] B.B. Straumal, A.A. Mazilkin, B. Baretzky, G. Schütz, E. Rabkin, R.Z. Valiev, Accelerated Diffusion and Phase Transformations in Co–Cu Alloys Driven by the Severe Plastic Deformation, Mater. Trans. 53 (2012) 63–71. https://doi.org/10.2320/matertrans.MD201111.

[52] R.G. Pirich, Gravitationally Induced Convection During Directional Solidification of off-Eutectic Mn-Bi Alloys, MRS Proc. 9 (1981) 593. https://doi.org/10.1557/PROC-9-593.

[53] H. Okamoto, Supplemental Literature Review of Binary Phase Diagrams: Ag-Co, Ag-Er, Ag-Pd, B-Ce, Bi-La, Bi-Mn, Cu-Ge, Cu-Tm, Er-Y, Gd-Tl, H-La, and Hg-Te, J. Phase Equilib. Diffus. 36 (2015) 10–21. https://doi.org/10.1007/s11669-014-0341-7.

[54] Yu.R. Kolobov, G.P. Grabovetskaya, M.B. Ivanov, A.P. Zhilyaev, R.Z. Valiev, Grain boundary diffusion characteristics of nanostructured nickel, Scripta Materialia. 44 (2001) 873–878. https://doi.org/10.1016/S1359-6462(00)00699-0.

[55] G. Wilde, S. Divinski, Grain Boundaries and Diffusion Phenomena in Severely Deformed Materials, Mater. Trans. 60 (2019) 1302–1315. https://doi.org/10.2320/matertrans.MF201934.

[56] M.D. Baró, Y.R. Kolobov, I.A. Ovidko, R.Z. Valiev, I.V. Alexandrov, M. Ivanov, K. Reimann, A.B. Reizis, S. Suriñash, A.P. Zhilyaev, DIFFUSION AND RELATED PHENOMENA IN BULK NANOSTRUCTURED MATERIALS, (n.d.) 43.

[57] O.L. Boothby, D.H. Wenny, E.E. Thomas, Recrystallization of MnBi Induced by a Magnetic Field, Journal of Applied Physics. 29 (1958) 353–353. https://doi.org/10.1063/1.1723130.

[58] Y. Mitsui, R.Y. Umetsu, K. Koyama, K. Watanabe, Magnetic-field-induced enhancement for synthesizing ferromagnetic MnBi phase by solid-state reaction sintering, Journal of Alloys and Compounds. 615 (2014) 131–134. https://doi.org/10.1016/j.jallcom.2014.06.131.

[59] H. Yasuda, I. Ohnaka, Y. Yamamoto, K. Tokieda, K. Kishio, Formation of Crystallographically Aligned BiMn Grains by Semi-solid Processing of Rapidly Solidified Bi-Mn Alloys under a Magnetic Field, Mater. Trans. 44 (2003) 2207–2212. https://doi.org/10.2320/matertrans.44.2207.

[60] T. Chen, W.E. Stutius, Magnetic-field-induced recrystallization in MnBi, Journal of Applied Physics. 45 (1974) 4622–4625. https://doi.org/10.1063/1.1663100.

[61] W.Y. Zhang, P. Kharel, T. George, X.Z. Li, P. Mukherjee, S. Valloppilly, D.J. Sellmyer, Grain alignment due to magnetic-field annealing in MnBi:Bi nanocomposites, J. Phys. D: Appl. Phys. 49 (2016) 455002. https://doi.org/10.1088/0022-3727/49/45/455002.

[62] G. Gottstein, Materialwissenschaft und Werkstofftechnik, Springer, Berlin, Heidelberg, 2014. https://doi.org/10.1007/978-3-642-36603-1.

[63] K. Kobayashi, R. Skomski, J.M.D. Coey, Dependence of coercivity on particle size in $Sm_zFe_{17}N_3$ powders, Journal of Alloys and Compounds. (1995).

[64] V. Ly, X. Wu, L. Smillie, T. Shoji, A. Kato, A. Manabe, K. Suzuki, Low-temperature phase MnBi compound: A potential candidate for rare-earth free permanent magnets, Journal of Alloys and Compounds. 615 (2014) S285–S290. https://doi.org/10.1016/j.jallcom.2014.01.120.

[65] K. Suzuki, X. Wu, V. Ly, T. Shoji, A. Kato, A. Manabe, Spin reorientation transition and hard magnetic properties of MnBi intermetallic compound, Journal of Applied Physics. 111 (2012) 07E303. https://doi.org/10.1063/1.3670505.

[66] J. Borsup, T. Eknapakul, H. Thazin Myint, M.F. Smith, V. Yordsri, S. Pinitsoontorn, C. Thanachayanont, T. Zaw Oo, P. Songsiriritthigul, Formation and magnetic properties of low-temperature phase manganese bismuth prepared by low-temperature liquid phase sintering in vacuum, Journal of Magnetism and Magnetic Materials. 544 (2022) 168661. https://doi.org/10.1016/j.jmmm.2021.168661.

[67] W.K. Unger, M. Stolz, Growth of MnBi Films on Mica, (1971).

[68] S. Suwas, S. Mondal, Texture Evolution in Severe Plastic Deformation Processes, Mater. Trans. 60 (2019) 1457–1471. https://doi.org/10.2320/matertrans.MF201933.

[69] F. Staab, E. Bruder, L. Schäfer, K. Skokov, D. Koch, B. Zingsem, E. Adabifiroozjaei, L. Molina-Luna, O. Gutfleisch, K. Durst, Hard Magnetic $SmCo_5$-Cu Nanocomposites Produced by Severe Plastic Deformation, (2022). https://doi.org/10.2139/ssrn.4263460.

[70] K. Kang, L.H. Lewis, A.R. Moodenbaugh, Alignment and analyses of MnBi∕Bi nanostructures, Appl. Phys. Lett. 87 (2005) 062505. https://doi.org/10.1063/1.2008368.

[71] Z. Ren, X. Li, H. Wang, K. Deng, Y. Zhuang, The segregated structure of MnBi in Bi–Mn alloy solidified under a high magnetic field, Materials Letters. 58 (2004) 3405–3409. https://doi.org/10.1016/j.matlet.2004.04.021.

[72] P.Z. Si, Y. Yang, L.L. Yao, H.D. Qian, H.L. Ge, J. Park, K.C. Chung, C.J. Choi, Magnetic-field-enhanced reactive synthesis of MnBi from Mn nanoparticles, Journal of Magnetism and Magnetic Materials. 476 (2019) 243–247. https://doi.org/10.1016/j.jmmm.2018.12.077.

[73]  W.K. Unger, H. Harms, Grain size of MnBi films, Appl. Phys. 2 (1973) 191–196. https://doi.org/10.1007/BF00884210.

[74]  W.P. Ellis, N.H. Nachtrieb, Anomalous Self-Diffusion in Solid Bismuth: The Trapping Mechanism, Journal of Applied Physics. 40 (1969) 472–476. https://doi.org/10.1063/1.1657422.

[75]  Y. Yanaka, Y. Kariya, H. Watanabe, H. Hokazono, Plastic Deformation Behavior and Mechanism of Bismuth Single Crystals in Principal Axes, Mater. Trans. 57 (2016) 819–823. https://doi.org/10.2320/matertrans.MD201503.

[76]  A.L. Ruoff, R.W. Balluffi, Strain-Enhanced Diffusion in Metals. II. Dislocation and Grain-Boundary Short-Circuiting Models, Journal of Applied Physics. 34 (1963) 1848–1853. https://doi.org/10.1063/1.1729698.

**Supplementary**

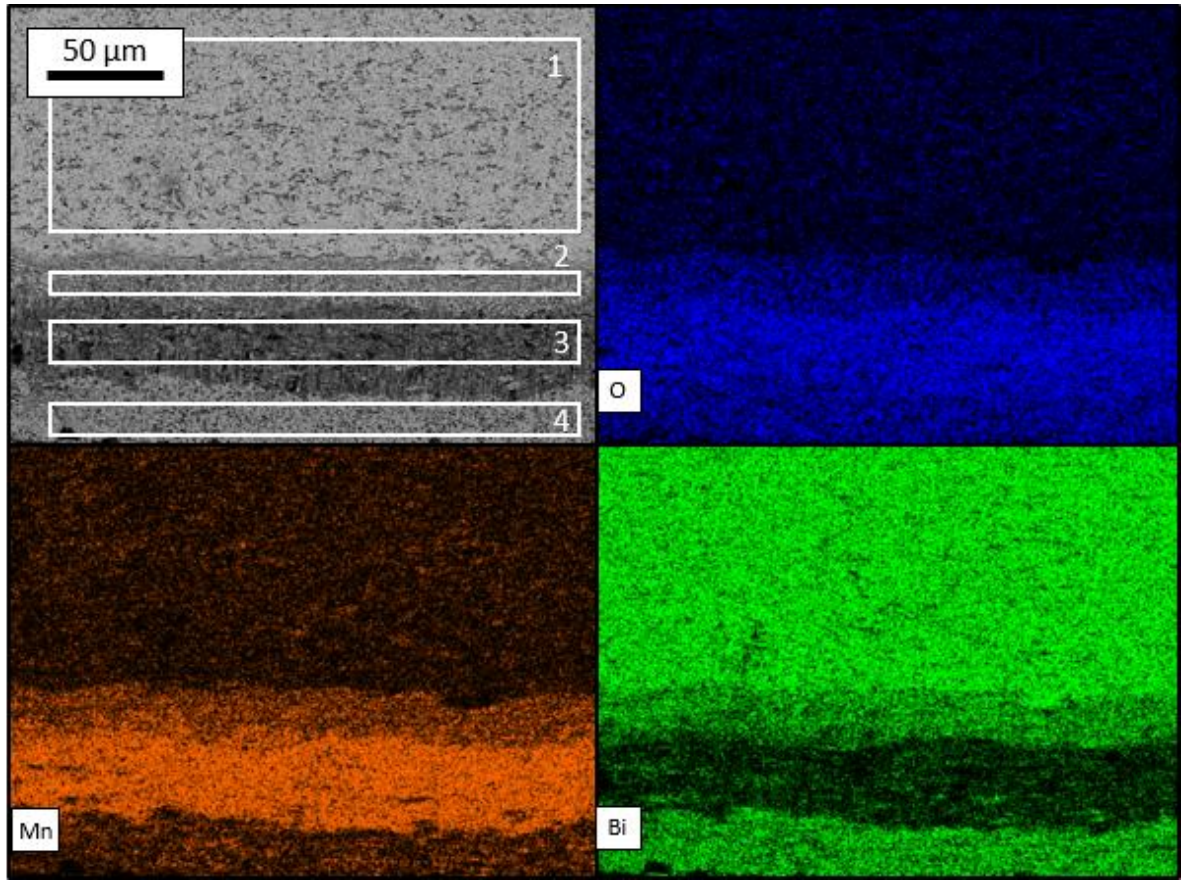

Figure Supplementary 1: Secondary electron image and corresponding energy dispersive X-ray spectroscopy mapping analysis of an MnBi composite after 120 h conventional annealing at 230°C. The images are recorded in tangential HPT-disc direction on the HPT-disc edge with the surface normal (axial HPT-disc direction) pointing downwards. The color intensity of each graph is correlated to their relative occurrence on the map. The chemical compositions for the marked regions and the entire map are summarized in Table Supplementary 1.

Table Supplementary 1: Chemical composition obtained from EDX analysis.

|  | Mn [at. %] | Bi [at. %] | O [at. %] |
|---|---|---|---|
| total map | 29,4 | 24,3 | 46,3 |
| area 1 | 24,1 | 46,1 | 29,8 |
| area 2 | 34,6 | 18,4 | 47,0 |
| area 3 | 41,3 | 6,5 | 52,3 |
| area 4 | 21,6 | 18,9 | 59,5 |

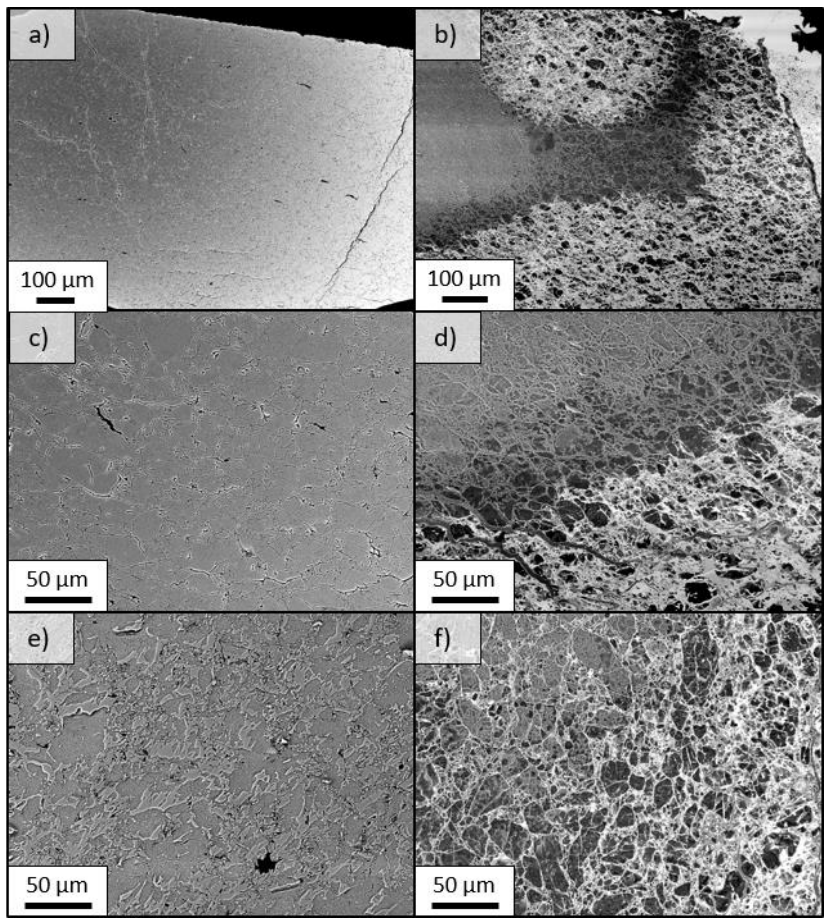

Figure Supplementary 2: BSE images of an HPT-deformed a) and c) and an HPT-compacted e) sample. Without any further pretreatment, the same samples were immersed into molten Bi at 410°C for 4 h b), d) and f). The Bi phase diffuses inside the bulk material, along the Mn grain boundaries for several 100 µm.